\documentclass[a4paper,11pt]{article}
\pdfoutput=1 
\usepackage{jcappub}
\usepackage[utf8]{inputenc}
\usepackage{subfigure}
\usepackage{epsfig} 
\usepackage{amsmath} 
\usepackage{amsfonts}  
\usepackage{float}    
\usepackage{amssymb}   
\usepackage{slashed} 
\usepackage{color}       
\usepackage{pbox}
\usepackage{array,multirow,makecell}
\usepackage{natbib} 
\usepackage{lipsum}
\usepackage{appendix}

\newcommand{\lsim}[1]{\lesssim}
\usepackage[normalem]{ulem}

\makeatletter
\gdef\@fpheader{}
\makeatother

\usepackage{orcidlink}
\definecolor{lime}{HTML}{A6CE39}
\DeclareRobustCommand{\orcidicon}{
	\begin{tikzpicture}
	\draw[lime, fill=lime] (0,0) 
	circle [radius=0.2] 
	node[white] {{\fontfamily{qag}\selectfont \tiny ID}};
	\draw[white, fill=white] (-0.0625,0.095) 
	circle [radius=0.007];
	\end{tikzpicture}
	\hspace{-2mm}
}

\foreach \x in {A, ..., Z}{\expandafter\xdef\csname orcid\x\endcsname{\noexpand\href{https://orcid.org/\csname orcidauthor\x\endcsname}
			{\noexpand\orcidicon}}
}

\newcommand{\be}{\begin{equation}}
\newcommand{\ee}{\end{equation}}
\newcommand{\bea}{\begin{eqnarray}}
\newcommand{\eea}{\end{eqnarray}}

\def\dr{\text{DR}}

\def\bsm{\text{BSM}}

\newcommand{\nfc}{N_{\rm eff}^{\rm CMB}}

\newcommand{\lt}{\left }
\newcommand{\rt}{\right }

\newcommand{\Planck}{\textit{Planck}}

\begin{document}
\title{CMB signatures of gravity-mediated dark radiation in $\mathbf{\Delta N_{\rm eff}}$}

\author[a]{Anish Ghoshal\orcidlink{}}
\emailAdd{anish.ghoshal@fuw.edu.pl}
\affiliation[a]{
Department of Physics and Astronomy, University of Sussex, \\
Brighton, BN1 9RH, United Kingdom}

\author[b,c]{Sk Jeesun~\orcidlink{0009-0005-2344-9286},}
\emailAdd{jeesun@sjtu.edu.cn}
\affiliation[b]{State Key Laboratory of Dark Matter Physics, Tsung-Dao Lee Institute \& School of Physics and Astronomy, Shanghai Jiao Tong University, Shanghai 200240, China}
\affiliation[c]{Key Laboratory for Particle Astrophysics and Cosmology (MOE) \& Shanghai Key Laboratory for Particle Physics and Cosmology, Shanghai Jiao Tong University, Shanghai 200240, China}

\author[d,e,f,g]{Kazunori Kohri}
\emailAdd{kazunori.kohri@gmail.com}
\affiliation[d]{Division of Science, National Astronomical Observatory of Japan, Mitaka, Tokyo 181-8588, Japan}
\affiliation[e]{The Graduate University for Advanced Studies (SOKENDAI), Mitaka, Tokyo 181-8588,
Japan}
\affiliation[f]{Theory Center, IPNS, KEK, 1-1 Oho, Tsukuba, Ibaraki 305-0801, Japan}
\affiliation[g]{Kavli IPMU (WPI), UTIAS, The University of Tokyo, Kashiwa, Chiba 277-8583, Japan.}

\abstract{ 
Measurement of $N_{\rm eff}$ in the CMB (Cosmic Microwave Background) observations, like \Planck- 2018 and BBN (Big Bang Nucleosynthesis) has already set stringent constraints on the interaction strength of light particles beyond the Standard Model (BSM). Despite such negligible couplings of such BSM particles to the visible sector, they are inevitably produced in the early universe through gravity-mediated processes. If a sizable density of light particles survives around CMB formation, they may act as dark radiation contributing to $N_{\rm eff}$ at CMB epoch.
In this work, we study the production of such light \bsm~  particles through the gravity-mediated scatterings in an effective field theory (EFT) setup 
assuming that all non-gravitational couplings of the \bsm\ particle are negligible. 
Since the production is sensitive to the spin of the produced particle, we perform a concrete analysis for two representative cases: scalar dark Higgs \dr~ and vector dark photons \dr.
Using the \Planck-2018 observations we find constraints on the reheating temperature ($T_{\rm RH}$) and background equation of state ($w_\Phi$) during reheating in such scenarios featuring dark Higgs and dark photon. A comparative discussion involving gravity-mediated production of Dirac right-handed neutrinos ($\nu_R$) and light axion-like particles (ALP) is also presented.
Finally, for completeness, we also analyze the scenario where the production occurs through a generic spin-2 mediator characterized by an effective scale $\Lambda$ delineating the parameter space that is currently ruled out from Planck-2018 and can be probed by the future CMB experiments like LiteBird, Simon Observatory, CMB-S4, CMB-HD.
}
\maketitle

\section{Introduction}
The Cosmic Microwave Background (CMB) radiation originated in the early universe at a redshift of 
$z\lesssim1100$, offers crucial information about the the early cosmos \cite{Kolb:1990vq,Dodelson:2003ft}.
Since the epoch of recombination, when the hot, dense universe cooled sufficiently for neutral hydrogen atoms to form and became transparent, CMB photons have been freely propagating through the universe.
Various observables deduced from such CMB photon survey can thus reveal the thermal history of the universe.
$\nfc$ is one of such important cosmological parameters 
to shape our understanding of particle phenomena in the early epoch, which quantifies the effective number of relativistic degrees of freedom in the CMB epoch. At very high temperatures ($T\gg \mathcal{O}$(MeV)), photons (coupled to electrons through QED strength) and neutrinos were in thermal equilibrium via the weak interactions. This interaction rate eventually fell below the Hubble expansion rate due to the Universe's cooling, which led to the decoupling of the neutrino ($\nu_L$) from the photon ($\gamma$) bath \cite{Kolb:1990vq,Dodelson:2003ft}.
The difference in the energy densities of neutrino and photon baths is encoded through their ratio parametrized by $\nfc$.
Within the Standard Model (SM) scenario, this framework predicts a value of $N_{\rm eff}^{\rm SM}=3.046$ rather than exactly 3 \cite{Mangano:2005cc,Grohs:2015tfy,EscuderoAbenza:2020cmq}. 
The slight excess over $3$ corresponding to the number of active neutrino species in the SM arises from several effects like non-instantaneous neutrino decoupling, finite-temperature QED corrections, and neutrino flavour oscillations \cite{Mangano:2005cc,Grohs:2015tfy}.
CMB observations e.g. Planck 2018 can also independently measure the $\nfc$ which reports a precise measurement with  $95\%$ confidence level, $\nfc = 2.99^{+0.34}_{-0.33}$ \cite{Planck:2018vyg}. 
The constraint remains mostly consistent with the SM prediction.

On the other hand, $\nfc$ is extremely sensitive to the presence of any kind of extra radiation \cite{Abazajian:2019oqj,Poulin:2018cxd,Ghosh:2022fws,Ghosh:2023ocl}, BSM light mediators interacting between $e,\gamma$ or $\nu_L$ \cite{Escudero:2018mvt,Escudero:2019gzq,Li:2023puz,Esseili:2023ldf,Ghosh:2023ilw,Ghosh:2024cxi}, non-standard neutrino interactions (NSI) \cite{Biswas:2022fga,Berbig:2020wve} etc. 
The Planck 2018 limit on $\nfc$ can be used to constrain several BSM scenarios featuring dark radiation or light mediators and already have placed stringent constraints on the SM interactions of such light BSM particles \cite{Escudero:2019gzq,Li:2023puz,Esseili:2023ldf,Ghosh:2023ilw,Ghosh:2024cxi,Ghosh:2022fws,Ghosh:2023ocl}, or couplings of the inflaton sector to discriminate between inflation models \cite{Ghoshal:2023phi}.
Even if the SM interactions (e.g. Yukawa, gauge) of the BSM species are negligible, weaker gravity mediated interactions with SM fields are always active \cite{Ema:2015dka,Garny:2015sjg}. 
The presence of the irreducible gravity-mediated production can thus populate the light BSM species in the early universe. 
 If such particles are produced with a sufficiently large energy density, they may persist until the epoch of CMB formation, where they would behave as dark radiation and contribute to $N_{\rm eff}$. Consequently, observations from Planck 2018 \cite{Planck:2018vyg} will impose stringent constraints on these scenarios.
Apart from that, there also exist constraint from ACT 2025 data which, combined with Planck LB leads to  $\nfc = 2.86^{+0.13}_{-0.13}$ at $68\%$ C.L., \cite{AtacamaCosmologyTelescope:2025nti}.
Upcoming CMB experiments, such as LiteBird ($\nfc\lesssim 3.19$) \cite{LiteBIRD:2022cnt}, Euclid ($\nfc\lesssim 3.11$ at $95\%$ C.L.) \cite{Euclid:2024imf}, Simons Observatory (SO) ($\nfc\lesssim 3.12$ at $95\%$ C.L.) \cite{SimonsObservatory:2019qwx}, SPT-3G ($\nfc\lesssim 3.09$ at $95\%$ C.L.)  \cite{SPT-3G:2024atg}, CMB-S4 ($\nfc\lesssim 3.10$ at $95\%$ C.L.) \cite{CMB-S4:2016ple},  PICO ($\nfc\lesssim 3.10$ at $95\%$ C.L.) \cite{NASAPICO:2019thw}, CMB-HD ($\nfc\lesssim3.06$) \cite{Sehgal:2019ewc} are expected to achieve the sensitivity to probe even lower values of $\nfc$.
The cosmic variance limited (CVL) is expected to restrict $\Delta N_{\rm eff}=\nfc-N_{\rm eff}^{\rm SM}\lesssim 3.1\times 10^{-6}$ \cite{Ben-Dayan:2019gll}.
Hence, the gravitationally produced DR can also be traced through its imprints in $\nfc$, which stands as the primary goal of this work.

The gravity mediated interactions can be realized in an effective field theory (EFT) framework where the spin-2 graviton couples to the stress-energy momentum tensor of the particles with the ultra-violet (UV) scale fixed to be the Planck mass, set from the equivalence principle \cite{Ema:2015dka,Garny:2015sjg,Garny:2017kha,Tang:2016vch,Tang:2017hvq,Ema:2016hlw,Bernal:2018qlk}.
Through the gravity-mediated SM scattering BSM particles can be produced in the early universe \cite{Ema:2015dka,Garny:2015sjg,Garny:2017kha,Tang:2016vch,Tang:2017hvq,Ema:2016hlw,Bernal:2018qlk,Ema:2018ucl,Ema:2019yrd,Chianese:2020yjo,Chianese:2020khl,Redi:2020ffc,Mambrini:2021zpp,Barman:2021ugy,Haque:2021mab,Clery:2021bwz,Clery:2022wib,Ahmed:2022qeh,Ahmed:2022tfm} and such production depends on the reheating temperature ($T_{RH}$) of the universe as they occur through operators with dimension, $D>4$ \cite{Elahi:2014fsa}. 
Besides, the gravity-mediated production from SM bath, contribution from the inflaton scattering somtimes become crucial, see Refs. \cite{Mambrini:2021zpp,Barman:2021ugy,Haque:2021mab,Clery:2021bwz,Clery:2022wib,Haque:2023yra,Barman:2022qgt,Bernal:2021kaj}.
Such studies exploring the gravity-mediated production of particles during the process of reheating have analyzed dark matter (DM) and matter-antimatter asymmetry production involving dark sector particles \cite{Garny:2015sjg,Tang:2016vch,Bernal:2018qlk,Ema:2019yrd,Chianese:2020yjo,Ahmed:2020fhc,Kolb:2020fwh,Redi:2020ffc,Ling:2021zlj,Bernal:2021kaj,Barman:2021ugy,Haque:2021mab,Clery:2022wib,Co:2022bgh,Garcia:2022vwm,Kaneta:2022gug,Mambrini:2022uol,Basso:2022tpd,Barman:2022qgt,Haque:2023yra,Kolb:2023dzp,Garcia:2023awt,Kaneta:2023kfv,Garcia:2023qab,Garcia:2023dyf,Garcia:2023obw,Kolb:2023ydq,Clery:2024dlk,Racco:2024aac,Dorsch:2024nan,Choi:2024bdn,Verner:2024agh,Jenks:2024fiu,Mondal:2025kur}.
Apart from these, gravity-mediated production can also generate a significant population of dark radiation (DR) as hinted before.
The \dr~ can be in the form of light sterile neutrino~\cite{Archidiacono:2016kkh, Archidiacono:2014apa}, neutralino~\cite{Bae:2013hma}, dark photon \cite{Fabbrichesi:2020wbt},  axions~\cite{DiValentino:2013qma,DEramo:2018vss}, Goldstone bosons~\cite{Weinberg:2013kea}, early dark energy~\cite{Calabrese:2011hg}.
On the other hand, if such \dr~ possess negligible non-gravitational interaction with SM fields, they remain mostly invisible to laboratory searches.

In this analysis, we explore this alternate possibility that DR is produced inevitably via gravity-mediated production, and  then it may leave its imprints on \dr~ measurements as $\Delta N_{\rm eff}$ in CMB. 
As highlighted before, such production is dominated around the reheating temperature. To incorporate the dynamics during the reheating epoch in a generic and minimal scenario, we consider a scalar field $\Phi$ whose energy density dominates the energy budget during the reheating epoch. However, for convenience, throughout this paper, we refer to this generic field as ``inflaton" \footnote{\emph{Although we denote the scalar $\Phi$ as the inflaton, what we consider is essentially any spectator field that is active and dominates the energy budget of the Universe during the reheating era. For the actual inflaton, a detailed inflationary model-dependent analysis CMB analysis will be needed, see Ref.\cite{Ghoshal:2025ejg} which is beyond the scope of the current paper. Several well-motivated examples of such spectator field responsible for reheating involves the curvaton, reheaton, Affleck-Dine, modulated reheating, inhomogeneous end of reheating and instant preheating scenarios \cite{Felder:1998vq,Dimopoulos:2017tud,Dimopoulos:2018wfg,Opferkuch:2019zbd,Bettoni:2021zhq,Laverda:2023uqv,Figueroa:2024asq,Ghoshal:2022ruy,Ghoshal:2024wom}.}}.
Our generic result can be easily extended to other BSM scenarios as well to realistic inflaton models, though not restricted to the case of the inflaton only.
In this spirit, we consider the minimal setup with SM particle content augmented by the inflaton $\Phi$ and dark radiation $X$.
We also assume negligible other couplings of BSM species so that gravitational production is the only source for their density.
In such setup, we study particle production in the processes (1) $\Phi\Phi\to$DR DR and (2) $\Phi\Phi\to$SM SM  with consecutive SM SM$\to$DR DR  via the s-channel graviton exchange.
The production depends on the spin of $X$ and for example we consider two scenarios: (1) scalar \dr~ ($S$) and (2) vector \dr~ ($A'$).
Since the process is sensitive to the reheating temperature ($T_{\rm RH}$), we find constraints on the $T_{\rm RH}$ from the contribution in $\nfc$.
Alongside the inflaton equation of state ($w_{ \Phi}$) is constrained from current CMB measurements.
The results with other BSM particles like axion like particles or light fermions can also be apprehended from our analysis.
Finally for generalization, we also study the \dr~ production through a generic spin-2 mediator exchange with the EFT scale being the free parameter.  
Future CMB missions will be able to probe quite a large region of parameter space of such gravity-mediated DR production, which is otherwise difficult to probe in ground-based experiments due to their negligible coupling.

\textit{The paper is organized as follows:} we describe the gravity-mediated production of light particles as dark radiation in Sec.\ref{sec:prod}. In Sec.\ref{sec:scalar} we study the production of Higgs as dark radiation and in  Sec.\ref{sec:vector} we investigate the same for a dark photon and in  Sec.\ref{sec:genspin2} we look at the production of light particles mediated via generic spin-2 mediators. 
We briefly comment on the production other BSM species through gravity-mediated processes in  Sec.\ref{sec:ALP}.
We end with conclusions and future outlook in Sec.\ref{sec:concl}.

\section{Gravity-mediated production of dark radiation}
\label{sec:prod}

\begin{figure}[!tbh]
    \centering
    \includegraphics[scale=0.15]{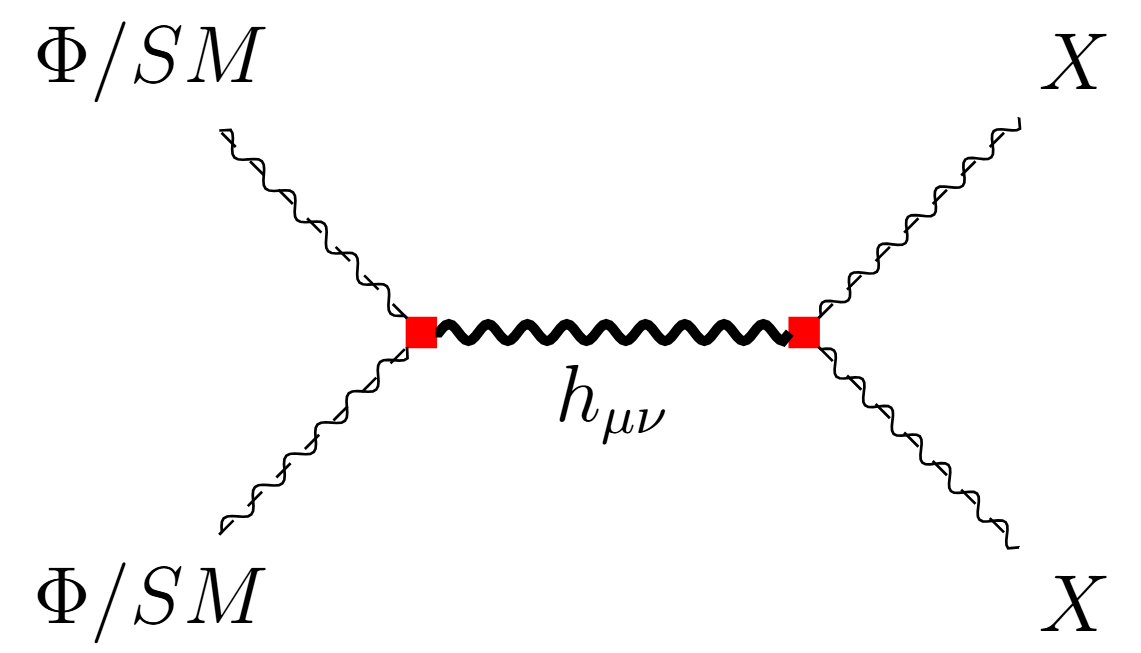}
    \caption{\it X as DR production from inflaton $\Phi$ or SM particles via graviton exchange or gravity-mediated process.}
    \label{fig:fd}
\end{figure}

We embark on a minimal scenario with an additional BSM particle $X$ singlet under SM gauge group, apart from the SM particles and the inflaton field $\Phi$ with the following interaction \textit{Lagrangian} \cite{Garny:2017kha,Bernal:2018qlk}, 
\be
\sqrt{-g}\mathcal{L}_{\rm int}= \frac{1}{2 M_{pl}} h_{\mu \nu}\lt(T^{\mu \nu}_{\rm SM}+ T^{\mu \nu}_{X}+ T^{\mu \nu}_{\Phi}\rt),
\label{eq:gravint}
\ee
where, $T^{\mu \nu}_i,~{i\equiv {\rm SM},\Phi,X}$ is the stress energy momentum tensor. $h_{\mu \nu}$ and $M_{pl}$ signifies the massless graviton field and the reduced planck mass scale ($2.4 \times 10^{18}$ GeV) respectively. The above eq.\eqref{eq:gravint} is obtained after expanding the metric around the flat Minkowski space-time ($\eta_{\mu \nu}$) i.e. $g_{\mu \nu}=\eta_{\mu \nu}+\frac{h_{\mu \nu}}{2 M_{pl}}$.
Note that the gravitational interaction depend mostly on the energy momentum tensor which in turn depends on the spin of the field.
Depending upon the spin of the particle, $T^{\mu \nu}_s$ for spin $s=0,1/2,1$ reads as:
\bea
T^{\mu \nu}_0&=& \partial^\mu \phi \partial^\nu \phi -g^{\mu \nu} \lt( \frac{1}{2}\partial^\lambda \phi \partial_\lambda \phi -V(\phi)\rt) \label{eq:ten1}\\
T^{\mu \nu}_{1/2}&=& \frac{i}{4}\bigg(\Bar{f} \gamma^\mu\partial^\nu f +\Bar{f} \gamma^\nu\partial^\mu f \bigg) -g^{\mu \nu} \lt( \frac{i}{2}\Bar{f}\gamma^\lambda\partial_\lambda f -m_f \Bar{f}{f}\rt)\\
T^{\mu \nu}_{1}&=& \frac{1}{2}\lt(V^\mu_\lambda V^{\nu \lambda} +V^\nu_\lambda V^{\mu \lambda}  - \frac{1}{2}g^{\mu \nu}V^{\lambda \delta}V_{\lambda \delta} \rt)
\label{eq:ten3},
\eea
where, $\phi,f$ represent any generic scalar and fermion respectively. $V^{\mu \nu}=\partial^\mu A^\nu-\partial^\nu A^\mu$ signifies field tensor of any vector field $A^\mu$.
Through graviton $h_{\mu \nu}$ mediated process both SM and $X$ particle can be produced (see Fig.\ref{fig:fd}).
As mentioned earlier, the production of $X$ will strongly depend on its spin as evident from the aforementioned eq.\eqref{eq:ten1}-eq.\eqref{eq:ten3}.
It is also worth highlighting that the gravity-mediated processes do not depend on the ``particle" coupling (e.g. strong or electroweak interactions) of the involved particle, either \cite{Bernal:2018qlk,Barman:2022qgt}.
\emph{Thus, for any BSM particle, even with negligible coupling strength with SM sector, they can be inevitably produced in the early universe.}
In the following subsection, we analyze the imprints of such BSM particles in the CMB through the contribution in $N_{\rm eff}$.

\subsection{Evolution of $N_{\rm eff}$}
\label{subsec:neff}
If BSM particles are light enough that, they may remain relativistic around the SM neutrino decoupling epoch ($\sim 2$ MeV) they may contribute to the extra radiation energy density and even might alter the neutrino decoupling or neutrino bath temperature (See Ref. \cite{Fields:2019pfx} for a review.) 
The presence of any such extra radiation is parametrized by the variable ``number of relativistic degrees of freedom" i.e. $N_{\rm eff}$.
At the time of CMB, this number of relativistic degrees of freedom is expressed as, 
\begin{equation}
N_{\rm eff}^{\rm CMB} = \frac{8}{7} \left(\frac{11}{4} \right)^{4/3} \frac{\rho_{rad}-\rho_\gamma}{\rho_{\gamma}}\Bigg|_{\rm T=T_{CMB}},
\end{equation}
where, $\rho_{\rm rad}$ and $\rho_{\gamma}=2 \times \frac{\pi^2}{30} T_\gamma^4 $ are energy densities of total radiation and photon  respectively. The photon bath temperature is denoted by $T_\gamma$. $T_{\rm CMB}$ stands for the temperature around CMB formation. 
In the SM case it is given as,
\begin{equation}
N_{\rm eff}^{\rm CMB} = 
\frac{8}{7} \left(\frac{11}{4} \right)^{4/3} \frac{\rho_{\nu_L}}{\rho_{\gamma}}\Bigg|_{\rm T=T_{CMB}},
\end{equation}
where, $\rho_{\nu}^{\rm SM}=3 \times 2\times \frac{7}{8} \times \frac{\pi^2}{30} (T_{\nu})^4$ is the neutrino bath ($\nu_L$) energy density with $T_{\nu}$ 
being the neutrino bath temperature. 
The current and future CMB experiments are extremely sensitive to $\nfc$ \cite{Planck:2018vyg,CMB-S4:2016ple,CMB-HD:2022bsz,SimonsObservatory:2018koc}.
Therefore, the couplings with SM particles of any BSM light species are stringently constrained or can be tested in the near future through present and future generation CMB experiments.

\emph{However, even if the couplings of BSM species with SM particles are negligible, their production in the early universe through gravity-mediated processes is unavoidable.  
Any such light species produced with sufficient density in the early universe can affect $\nfc$.
Thus, such scenarios can be probed in present and future CMB experiments despite their negligible particle interaction strength with SM fields which stands as the main thrust of this analysis.} 
 In the presence of the BSM $X$ injection as a source of extra radiation density in our scenario, $N_{\rm eff}$ will be given by, 
\begin{equation}
  {N^{\prime} _{\rm eff}} ^{\rm CMB} = 
\frac{8}{7} \left(\frac{11}{4} \right)^{4/3} \frac{\rho_{\nu_L}+\rho_{X}}{\rho_{\gamma}}\Bigg|_{\rm T=T_{CMB}},  
\label{eq:neff}
\end{equation}
where, $\rho_X$ is the density of $X$.
In this case,
the relativistic neutrino degrees of freedom($N'_{\rm eff}$) also differs
from the prediction of SM at the time of CMB. 
For our case we assume $X$ is feebly interacting and does not participate in the interactions related to $\nu_L$ decoupling. The gravitational interaction of $X$ with $e,\nu_L$ is also significantly suppressed than compared to the weak interaction processes.
Hence, $X$ affects the $\nu_L$ decoupling through its density, contributing to the expansion rate.
Note that, this $X$ particle is often dubbed as dark radiation as mentioned in the Introduction.
Since the production occurs through dimensionful operators, the produced density depends on $T_{RH}$, which is more of a generic feature in UV freeze-in \cite{Elahi:2014fsa} and will be shown in detail in later sections (Sec.\ref{sec:scalar}-Sec.\ref{sec:genspin2}).
Once produced, $\rho_{X}$ dilutes only through redshift governed by the expansion term in the absence of any other sizable interactions. To track the $N_{\rm eff}$ we solve the following energy density equations,

\begin{eqnarray}
 \frac{d\rho_{\gamma}^{\rm tot}}{dt}&=&  - 4 H \rho_{\gamma} - 3 H (\rho_{e}+P_e) +\frac{\delta \rho_{e\to \nu_L}}{\delta t}, \label{eq:tgamma}\\
 \frac{d\rho_{\nu_L}}{dt}&=& \left(- 4 H \rho_{\nu_L} +\frac{\delta \rho_{\nu_L \to e}}{\delta t}\right) \label{eq:tnu}\\
 \frac{d\rho_{X}}{dt}&=& \left(- 4 H \rho_{X} +\mathcal{C}[T]\right),
   \label{eq:tx}
\end{eqnarray}
where, $\rho_{\gamma}^{\rm tot}=\rho_{\gamma}+\rho_e$ represents the total energy of photon bath around $\sim \mathcal{O}(1)$ MeV scale temperature. $\nu_L$ signifies all 3 generation of neutrinos ($\nu_i$, where $i=e,\mu,\tau$) as a whole i.e $\rho_{\nu_L}=\sum_{i=e,\nu,\tau}\rho_{\nu_i}$.

$\mathcal{C}[T]$ is the collision term relevant for the species $X$ which we will evaluate as a function of temperature $T$ as will be shown later. Since, we will be dealing with very high temperature (T $\gg$  decoupling temperature of $\nu_L$ i.e. $ \mathcal{O}$(MeV) \cite{Escudero:2018mvt}), we should in principle track all the relevant particles' density. However, as already mentioned, we assume the dark particles $X$ to have negligible ``particle " interaction with SM particles. 
On the other hand, the only SM particles with sufficient energy density relevant for deciding $\nu_L$ decoupling are $\gamma$ and $e^\pm$, whereas the heavier SM particles get Boltzmann suppressed \cite {Escudero:2018mvt,Ghosh:2023ilw}.
In such a scenario, once produced $X$ density just redshifts, hence one can simply solve the aforementioned three equations eq.\eqref{eq:tx} with energy transferred (between $e$ and $\nu_L$ bath) only by SM processes to track the $\nu_L$ decoupling \footnote{We also ignore perturbations in the dark sector radiation fluid, which may have consequences in the dark radiation iscourvature, see Ref.\cite{Ghosh:2021axu}.}.  
However, despite of negligible interaction, the produced density of $X$ at the early universe comes into play through the Hubble expansion rate $H$. Around $\mathcal{O}(10)$ MeV temperature, $H$ is given by,
\begin{eqnarray}
    H= \sqrt{(\rho_{\gamma}+\rho_e+\rho_{\nu_L}+\rho_X)}/M_{pl}
\end{eqnarray}
The energy transfer rates through the SM processes are given by \cite{Luo:2020sho,Escudero:2018mvt}:
\begin{eqnarray}
   \left(\frac{\delta \rho_{\nu_e \to e}}{\delta t}\right)&=&\frac{G_F^2}{\pi^5} \left\{(1+4s_W^2+8s_W^4) F_{\rm MB}(T_{\gamma},T_{\nu_e})\right\} \label{eq:delnu1} \\
   \left(\frac{\delta \rho_{\nu_{\mu/\tau}\to e}}{\delta t}\right)&=&\frac{G_F^2}{\pi^5} \left\{(1-4s_W^2+8s_W^4) F_{\rm MB}(T_{\gamma},T_{\nu_\mu})\right\}
    \label{eq:delnu2}
\end{eqnarray}
where, $\frac{\delta \rho_{i\to j}}{\delta t}$ encodes the energy density transfer rate from $i$ bath to $j$. The energy transfer rate quoted in eq.\eqref{eq:tgamma}-\eqref{eq:tx} is given as $ \left(\frac{\delta \rho_{\nu_L \to e}}{\delta t}\right)=\sum_{i=e,\mu,\tau}\left(\frac{\delta \rho_{\nu_i\to e}}{\delta t}\right)$. $G_F$ is the 4 Fermi constant and $s_W= \sin \theta_W$ where $\theta_W$ is the Weinberg angle. Finally the function 
$F_{\rm MB}$ is given by \cite{Escudero:2018mvt},
\begin{eqnarray}
    F_{\rm MB}(T_1,T_2) = 32\, (T_1^9-T_2^9) + 56 \, T_1^4\,T_2^4 \, (T_1-T_2)\, .
\end{eqnarray}
Plugging these collision terms, we solve the aforementioned eq.\eqref{eq:tgamma}-eq.\eqref{eq:tx}.
For eq.\eqref{eq:tgamma} and eq.\eqref{eq:tnu} it is easier to solve by transforming the energy densities $\rho_{\gamma,\nu}$ as a function of temperatures $T_{\gamma,\nu}$.

\subsection{Production in early universe}
\label{subsec:prod}
In this subsection, we sketch out the production of DR  species $X$.
Since the spin of $X$ particle is crucial in their production in the early universe, we analyze  
various possible natures of $X$ depending upon their spins e.g.,
\begin{enumerate}
    \item Dark scalar: $S$ with spin $0$.
    \item Dark vector: $A'$ with spin $1$ e.g. dark photon, $B-L$ gauge bosons etc.
\end{enumerate}
Though specifically we do not present the analysis with dark fermions with spin $1/2$ e,g. milicharged fermion, Dirac neutrinos $\nu_R$ etc, we can easily infer about them from our analysis.

To analyze the production of DR we need to incorporate the cosmic evolution during the reheating epoch. As hinted in the introduction, for a generic  analysis at that epoch, we consider a spectator field $\Phi$ with a potential dominating the energy budget that we refer as ``inflaton" for convenience \cite{Felder:1998vq,Dimopoulos:2017tud,Dimopoulos:2018wfg,Opferkuch:2019zbd,Bettoni:2021zhq,Laverda:2023uqv,Figueroa:2024asq,Ghoshal:2022ruy,Ghoshal:2024wom}.
For illustration we also assume  
quite a generic potential of $\Phi$ valid during the reheating era \footnote{This potential has been studied widely as what is known as $\alpha-$attractor in the literature for inflation.} (Refs.\cite{Kallosh:2013lkr,Kallosh:2013hoa,Kallosh:2013yoa,Kallosh:2013pby,Kallosh:2013maa,Galante:2014ifa,Kallosh:2022feu}), which the during the reheating epoch, can be expanded around the origin as \cite{Mambrini:2021zpp}
\be
V(\Phi)=\lambda \frac{\Phi^k }{M_{pl}^{k-4}}~~\lt({\rm for }~\Phi \ll M_{pl}\rt)
\label{eq:pot}
\ee
After inflation ends, the inflaton begins to oscillate about the minimum, and the
process of reheating begins.
While analysing the gravity-mediated production, we consider the two different possible scenarios:
           \begin{itemize}
           \item From inflaton field, of the form   $\Phi \Phi \rightarrow$ X X.
            \item From SM particles produced during reheating from the inflaton i.e. $\Phi \Phi \rightarrow$ SM SM$\to$ X X. 
            \end{itemize}
In the following sub-sections we discuss them one by one.



\subsubsection{From inflaton field}
\label{subsub:inf}
Since the inflaton also couples to gravity (see eq.\eqref{eq:gravint}), the inflaton itself can help to produce $X$ through graviton mediated processes.
Again, the produced density of the species $X$ from the inflaton will be governed by the Boltzmann equation \cite{Kolb:1990vq},
    \bea
    \frac{d\rho_{X}^\Phi}{dt} +4 H \rho_{X}^\Phi&=& \mathcal{C}_{\Phi\to X}[T],
    \label{eq:cact_inf}
    \eea
where, $\rho_{X}^\Phi$ stands for the energy density of $\Phi$ produced directly from inflaton scattering. $\mathcal{C}_{\Phi\to X}[T]$ is the collision term responsible for the production of $X$ from $2\to2$ scattering of $\Phi$ and is given by,
\bea
\mathcal{C}_{\Phi\to X}[T] &=& \prod_{j=\Phi,\Phi} d^3\Pi_j f_\Phi f_\Phi \prod_{i=3,4} d^3\Pi_i |\mathcal{M}|^2_{\Phi \Phi \to X X} E_X,
\label{eq:cphi}
\eea
where, $d\Pi_i$ are the Lorentz invariant phase spaces and $f_i$ are the respective distribution functions.
$|\mathcal{M}|^2_{\Phi \Phi \to X X}$ is the amplitude square for the $s$ channel process $\Phi \Phi \to X X$, whereas $E_X$ is the energy of the produced $X$.Throughout this study we work in FLRW (Friedmann–Lemaître–Robertson–Walker) cosmology which explains a homogenous, isotropic expanding universe where infinitesimal distance can be written as $ds^2=dt^2-a(t)^2 dl^2$ with $a(t)$ being the scale factor.
In terms of this scale factor eq.\eqref{eq:cphi} can be further simplified as 
 \bea 
\frac{d(\rho_{X}^\Phi a^4)}{da}&=& \frac{a^3}{H}~\mathcal{C}_{\Phi\to X}[T].
\label{eq:rhoinf}
\eea
The Hubble expansion rate$H(=\dot{a}/a,~\dot{a}=da/dt)$ can also be expressed as a function of $a$.
During the reheating, the energy density is dominated by the inflaton itself i.e. $H\approx\sqrt{\rho_\Phi/(3 M_{pl}^2)}$, and hence one can write,
\begin{eqnarray}
    \rho_\Phi (a)&=& \rho_{\rm end} \bigg(\frac{a}{a_{ end}}\bigg)^{-\frac{6k}{k+2}}, \label{eq:rhoap}\\
     H &\approx&\frac{\sqrt{\rho_{\rm end}}}{\sqrt{3}M_p}\bigg(\frac{a}{a_{ end}}\bigg)^{-\frac{3k}{k+2}},
     \label{eq:hub}
\end{eqnarray}
where, $\rho_\Phi$ is the energy density of $\Phi$. $a_{\rm end}$  and $\rho_{\rm end}$ signify the scale factor and the energy density at the onset of the reheating epoch. Typically, $\rho_{\rm end}$ sets the typical scale of inflation which can be expressed as $\rho_{\rm end}=3 M_p^2 H_{\rm inf}^2$, with $H_{\rm inf}$ signifying the Hubble scale after inflation \cite{Garcia:2020wiy}. As mentioned earlier, though the $\Phi$ field in our scenario is not necessarily an inflaton in the strict sense, we still use the upper limit set on $H_{\rm inf}~(\lesssim 8.5 \times 10^{13}$ GeV) by CMB \cite{Artymowski:2017pua}.

To proceed with Eq.~\eqref{eq:rhoinf}, the remaining task is to compute the amplitudes for the process $\Phi \Phi \to X X$, evaluated under the assumption that $X$ is produced from an inflaton condensate modeled as an oscillating field \cite{Clery:2021bwz}.
This oscillating nature of the inflaton field is encoded through the following parametrization \cite{Clery:2021bwz},
\bea
\Phi(t)=\Phi_0(t) P(t),
\eea
where, $\Phi_0$ signifies the time-dependent amplitude and $P(t)$ governs the periodicity of the oscillation. Then the potential in eq.\eqref{eq:pot} can be replaced and expanded as \cite{Clery:2021bwz},
\bea
V(\Phi)=V(\Phi_0) P^k(t)= V(\Phi_0) \sum_{n=-\infty}^{n=\infty} P_n^k e^{-i\omega t}
\eea
with the frequency $\omega$ given as,
\bea
\omega= M_\Phi \sqrt{\frac{\pi k}{2(k-1)}} \dfrac{\Gamma(\frac{1}{2}+\frac{1}{k})}{\Gamma(\frac{1}{k})}.
\eea
In such a framework the amplitude for the process $|\mathcal{M}|^2_{\Phi \Phi \to X X}$  can be calculated as the summation of the amplitudes due to  each mode i.e. $|\mathcal{M}|^2_{\Phi \Phi \to X X}=\sum_{m=0}^\infty |\mathcal{M}|^2_{\Phi,m}$ with $ |\mathcal{M}|_{\Phi,m}$ being the amplitude for $X$ production from $m-$th mode of $\Phi$ \cite{Clery:2021bwz}.
$ |\mathcal{M}|^2_{\Phi,m}$ can be obtained as
\begin{eqnarray}
    |\mathcal{M}|^2_{\Phi,m} = |\mathcal{M}^{0\to X}_m|^2\times \left(P^k_m\right)^2,
\end{eqnarray}
where $|\mathcal{M}^{0\to X}_m|^2$ is the amplitude square of the process where 2 spin 0 particles collide at center of mass energy $\sqrt{s}=E_m=m\omega$ to produce 2 $X$ particles. For notational convenience from now on we refer $\mathcal{M}^{0\to X}_m$ as $\mathcal{M}_m$.

Finally we simplify  the rate in eq.\eqref{eq:cphi}  in the non-relativistic limit (NR) limit of $\Phi$ ($T\ll M_\Phi$).
In our region of interest, for $ 2 m_X\ll \sqrt{s}$,  the energy density of the produced light particle $E_X=\sqrt{s}/2$ can be approximated as $E_X\approx M_\Phi$ in the NR limit \cite{Barman:2021ugy}.
Thus the collision term becomes 
\bea
\mathcal{C}_{\Phi\to X}[T] &=& n_{\Phi}^2 M_\Phi \frac{g_X}{l!}\sum_{n=1}  \frac{1}{32 \pi M_{\Phi}^2}|\mathcal{M}_n|^2 \sqrt{1-\frac{4 m_X^2}{s}}\bigg|_{s\to E_n^2} \\
&\approx& n_{\Phi}^2 \frac{g_X}{l!}\sum_{n=1}  \frac{1}{32 \pi M_{\Phi}}|\mathcal{M}_n|^2 
\label{eq:cinf}.
\eea
$n_\Phi=\rho_\Phi/M_\Phi$ signifies the number density of $\Phi$, whereas 
$l$ describes the number of identical particles in the final state. $g_X$ stands as the number of produced $X$ particles in a single reaction.
Plugging this in the aforementioned  eq.\eqref{eq:rhoinf}, we can solve $\rho_X^\Phi$ in terms of $T_{RH}$, which will specifically depend on the spin of $X$ and will be separately discussed for each model considered here in the later sections. 
For a realistic inflaton model, $M_\Phi$ is governed by the potential and thus fixed for a given $k$ value. However, for a generic spectator field $\Phi$ we treat its mass $M_\Phi$ as a free parameter independent of $k$.

\subsubsection{From SM particles produced during reheating from the inflaton}
\label{subsub:sm}
Besides the inflaton scattering, SM particles produced during reheating (from the inflaton) can also contribute to $X$ production.  
Note that here we assume the SM radiation with energy density $\rho_R$ to be produced simultaneously from inflaton scattering.
Though $\rho_R$ initially is small, SM particles can also scatter through $s$ channel gravity-mediated process and may potentially produce BSM particles despite the negligible interaction strength other than gravitaional one. 
Thus, one needs to track both $\rho_\Phi$ and $\rho_R$ in order to compute the SM contribution in $X$ density i.e. $\rho_X^{\rm SM}$.
The evolution of energy densities is given by solving the coupled equations, 
\begin{eqnarray}
   \frac{d\rho_{\Phi}}{dt} +3 H (1+w_\Phi) \rho_{\Phi} &=& -(1+w_\Phi)\rho_{\Phi},\\
   \frac{d\rho_{R}}{dt} +4 H \rho_{R} &=& +(1+w_\Phi)\rho_{\Phi},\label{eq:sm}\\
   \frac{d\rho_{X}^{\rm SM}}{dt} +4 H \rho_{X}^{\rm SM}&=& \mathcal{C}_{{\rm SM}\to X}[T],
   \label{eq:x}
\end{eqnarray}
where, $\rho_\phi, w_\phi$ are the energy density and equation of state ($w_\Phi\approx \frac{k-2}{k+2}$) \cite{Garcia:2020wiy}.  
As mentioned earlier, we assume that the energy density is dominated by the inflaton during reheating, and hence the Hubble rate is governed by only $\rho_\Phi$ as well (see eq.\eqref{eq:hub}). 
For this work we do not assume any BSM coupling between $\Phi$ and SM particles except the gravity mediated interaction (eq.\eqref{eq:gravint}).
In such scenario, inflaton scattering to Higgs scalars typically dominates the production rate during reheating \cite{Garcia:2020wiy}. Usually the inflaton dissipation rate is parametrized as $\Gamma_{\Phi}\propto \rho_{\Phi}^l$ \cite{Garcia:2020wiy}.
In our case, the effective quartic interaction between $\Phi$ and SM Higgs
graviton exchange which 
leads to $l=\frac{3}{2}-\frac{1}{k}$ \cite{Barman:2022qgt}.
Using this and eq.\eqref{eq:rhoap} in the aforementioned eq.\eqref{eq:sm}, one can express the SM radiation energy density as \cite{Garcia:2020wiy,Barman:2022qgt},
\begin{eqnarray}
    \rho_{R}(a) &=&  \rho_{RH}\bigg(\frac{a_{RH}}{a}\bigg)^4\Bigg[\dfrac{1-\big(\frac{a_{end}}{a}\big)^{\frac{8k-14}{k+2}}}{1-\big(\frac{a_{end}}{a_{RH}}\big)^{\frac{8k-14}{k+2}}} \Bigg],
    \label{eq:rhoR}
\end{eqnarray}
where, $a_{\rm RH}$ and $\rho_{\rm RH}$ are the scale factor and energy density at the end of reheating.
$a_{\rm RH}$ is defined at the epoch when $\rho_{\Phi}(a_{\rm RH})=\rho_R(a_{\rm RH})=\rho_{\rm RH}$.
On the other hand, $a$ indicates the scale factor between the two epochs: $a_{\rm end}\ll a\ll a_{\rm RH}$. 
One can then evaluate $\rho_X^{\rm SM}$ from the simplified form of eq.\eqref{eq:x} as,
\bea 
\frac{d(\rho_{X}^{\rm SM}a^4)}{da}&=& \frac{a^3}{H}~\mathcal{C}_{{\rm SM}\to X}[T].
\label{eq:rhoi}
\eea
The corresponding collision term governing the contribution from SM sector is given as \cite{Elahi:2014fsa},
\begin{eqnarray}
    \mathcal{C}_{\rm SM\to X}[T]&=&\sum_{\rm SM }\prod_{i=1}^4\int d\Pi_i~ E_1 (2\pi)^4 \delta^4 (p_1 + p_2 - p_3 -p_4)|\mathcal{M}|^2 \left[f_{\rm SM}  f_{\rm SM}  \right]\,\\
    &=&  \frac{ T}{512 \pi^6}\, 4\pi\, \sum_{\rm SM } \int_{4 M_{\chi}^2}^{\infty} ds \mathcal{P}_{XX}  \mathcal{P}_{\rm SM ~SM}\, |\mathcal{M}|^2 \,\left( \frac{\sqrt{s}}{2}\right) \frac{1}{\sqrt s} K_{1}\left( \frac{\sqrt s}{T} \right)
\label{eq:cf2}
\end{eqnarray}
with, $\mathcal{P}_{ij} $ defined as,
\begin{equation}
\mathcal{P}_{ij}= \frac{1}{2 \sqrt{s}} \sqrt{s-(m_i+m_j)^2} \sqrt{s-(m_i-m_j)^2}.
\end{equation}
$i,j$ represent the relevant particles with mass $m_{i,j}$.
$|\mathcal{M}|^2_{SM\to X}$ signify the amplitude squares for the scattering $SM SM \to h_{\mu \nu}\to X X$. 
The summation includes contributions from all SM particles: scalar ($s=0$), fermion ($s=1/2$), and vector bosons ($s=1$). 
Note that gravity-mediated production amplitudes contain dimensionful coupling, which essentially leads to a dependence on $T_{RH}$, just like in the case of UV freeze-in \cite{Elahi:2014fsa}. 
In the high temperature regime one can neglect the SM particles' masses and treat them as relativistic particles.
Hence, all the contributions from different SM species add up in eq.\eqref{eq:cf2} leading to,
\bea
\mathcal{C}_{\rm SM\to X} &=&\frac{ T}{512 \pi^6}\, 4\pi\, \int_{4 M_{\chi}^2}^{\infty} ds \mathcal{P}_{XX}  \mathcal{P}_{\rm SM ~SM}\, \overline{|\mathcal{M}|^2}_{\rm total} \,\left( \frac{\sqrt{s}}{2}\right) \frac{1}{\sqrt s} K_{1}\left( \frac{\sqrt s}{T} \right),\\
\label{eq:csmact}
{\rm where,}&&\overline{|\mathcal{M}|^2}_{\rm total}= 4\overline{|\mathcal{M}|^2}_{0X} +45\overline{|\mathcal{M}|^2}_{\frac{1}{2}X}+12\overline{|\mathcal{M}|^2}_{1X},
\eea
where $|\mathcal{M}|^2_{sX},s\equiv0,1/2,1$ signify the amplitude squares for $SM SM \to h_{\mu \nu}\to X X$ with SM particle being scalar, fermion and vector respectively. The prefactors in the above expression account for the number of degrees of freedom of SM bath particles in each spin category.
At high temperature  limit the collision term  becomes,
\begin{eqnarray}
    \mathcal{C}_{\rm SM\to X} &=& \beta_X \frac{T^9}{M_p^4}= \frac{\beta_X}{M_p^4} \bigg(\frac{\rho_{R}}{c_*} \bigg)^{9/4},
    \label{eq:ctr}
\end{eqnarray}
where, $\beta_X$ is  the numerical prefactor which can be evaluated after plugging the amplitude in eq.\eqref{eq:csmact} and performing the integration in terms of a dimensionless variable $s/T^2$. $\rho_R$ is the SM radiation energy density and $c_*=\frac{\pi^2}{30}g_*(T)$.
Using eq.\eqref{eq:rhoR} $\rho_R$ can be replaced as a function of scale factors and thus the appropriate collision term (from eq.\eqref{eq:ctr}) looks like, 
\begin{eqnarray}
    \mathcal{C}_{\rm SM\to X} [T] &\approx& \beta_X \frac{1}{M_p^4 c_*^{9/4}} \rho_{RH}^{9/4}  \bigg(\frac{a_{RH}}{a}\bigg)^9\Bigg[\dfrac{1-\big(\frac{a_{end}}{a}\big)^{\frac{8k-14}{k+2}}}{1-\big(\frac{a_{end}}{a_{RH}}\big)^{\frac{8k-14}{k+2}}} \Bigg]^{9/4}
    \label{eq:cSM}
\end{eqnarray}
We substitute this collision term in the previously mentioned eq.\eqref{eq:rhoi}  
and for the ease of computation, one can rearrange the same equation in terms of $r=a/a_{RH}$, which leads to,
\bea
\frac{d(\rho_{X}^{\rm SM}r^4)}{dr} &=& \beta_X \dfrac{\sqrt{3}~\rho_{RH}^{9/4}}{\sqrt{\rho_{\rm end}}~M_p^3~ c_*^{9/4}}~\Bigg[\dfrac{1}{1-r_{\rm RH}^{\frac{8k-14}{k+2}}}\Bigg]^{9/4}~
    ~ r_{\rm RH}^9~ r^{-\frac{3k+12}{k+2}}\big[1-r^{\frac{8k-14}{k+2}} \big]^{9/4},
        \label{eq:rho_SM}
\eea
where, we signify $r_{\rm RH}=a_{\rm RH}/a_{\rm end}$.
The initial condition of the above equation is $\rho_{X}=0$ at $r=1 ~(a=a_{\rm end})$. 
While solving the equation, we also replace $\sqrt{\rho_{\rm end}}=\sqrt{\rho_{\rm RH}}~(r_{\rm RH})^{3k/(k+2)}$.
For simplicity, this analysis is restricted in the regime where $r\gg1$ i.e., $a_{\rm RH}\gg a_{\rm end}$ \cite{Barman:2022qgt}.
From eq.\eqref{eq:rho_SM} it can be apprehended that the final density $\rho_X^{\rm SM}(a)$ will depend only on the ratio $a/a_{\rm RH}$ and $\rho_{\rm RH}$. We express both of them in terms of $T_{\rm RH}$ to find $\rho_X^{\rm SM}(T)$ and consider $T_{\rm RH}$ to be free parameter as also done in the previous subsection.
The same equation will govern the gravity-mediated production of different species, except $\beta_X$ will be different depending on the spin nature of $X$.

Finally we add up both the contribution from inflaton and from SM bath i.e. $\rho_X^{\rm tot}(T)=\rho_X^{\Phi}(T)+\rho_X^{\rm SM}(T)$.
After having a detailed discussion of \dr~ production and its consequences in $\nfc$, we are now set to explore minimal BSM scenarios for practical implementation.

\section{Dark Higgs as Dark Radiation}
\label{sec:scalar}
To analyze the \dr~ production through gravity-mediated processes, we explore the minimal scenario featuring a light BSM scalar.
We extend the SM particle content with a BSM scalar singlet field $S$ with the following \textit{Lagrangian}\cite{OConnell:2006rsp},
\bea
\mathcal{L}_S\supset \frac{1}{2}\partial_\mu S \partial^\mu S -\frac{1}{2}m_S^2 S^2,
\label{eq:sclag}
\eea
where $m_S$ is the mass of the BSM scalar.
From a model building perspective, such an extended scalar sector is often dubbed as ``dark higgs"\cite{OConnell:2006rsp,Feng:2017vli}. For \dr analysis, such a particle is required to be relativistic around CMB formation and hence, we consider $m_S\ll T_{\rm CMB}\sim 0.1$ eV. 
For the sake of simplicity and conservative estimates, we work in the scenario where all other couplings of $S$ (e.g. gauge, Yukawa, quartic couplings with SM higgs) with SM fields are negligible, so that gravity-mediated processes are the only source of production, as hinted previously. 
Within such framework, we now discuss the production of such dark higgs $S$  in the earlier-mentioned two scenarios i.e. (1) from inflaton $\Phi$ and (2) from SM sector.

The energy density of produced $S$ from inflaton scattering can is obtained from the methodology prescribed in Sec.\ref{subsub:inf}.
As mentioned earlier, the information of $S$ production from the inflaton enters through the collision term in eq.\eqref{eq:rhoinf}.
Plugging the amplitude $|\mathcal{M}|^2$ of the process $\Phi\Phi\to S S$ (see ref.\cite{Barman:2021ugy}) in eq.\eqref{eq:cinf} we obtain  the simplified collision term, 
\bea
 \mathcal{C}_{\Phi\to S}[T]
&\approx & \frac{g_S}{l!} \frac{M_\Phi \rho_{\Phi}^2}{M_p^4} \Sigma_k^0
~~~{\rm where,} ~\Sigma_k^0 = \frac{1}{512~\pi } \sum_{n=1} |\mathcal{P}^k_n|^2 ,
\label{eq:cscal}
\eea
where we have dropped the terms linear to $\mathcal{O}(m_S^2/M_\Phi^2)$ which is a realistic approximation for our chosen parameter range.
For a real (complex) scalar, the information of identical particles in the final state is incorporated through $l$ with $l=2$ ($l=1$).
On the other hand, $g_s=2$ signifies the number of produced $S$ in a single reaction.
To proceed with the energy density evolution equation (eq.\eqref{eq:rhoinf}), we replace the inflaton density in the collision term as \cite{Basso:2022tpd},
\bea
\rho_\Phi(a) &=&\rho_{\rm RH} \bigg(\frac{a_{\rm RH}}{a}\bigg)^{\frac{6k}{k+2}}
\eea
and substitute in eq.\eqref{eq:rhoinf} to obtain the energy density ($\rho_{S}^{\Phi}$) evolution in terms of the scale factor $a$ as,
\bea
\frac{d(\rho_{S}^{\Phi}(a) a^4)}{da}&=& \frac{\sqrt{3}}{M_p^3} \rho _{\text{RH}}^{3/2} 
   \left(a_{\rm RH}\right)^{\frac{9 k}{k+2}} \left(a \right)^{\frac{6-6 k}{k+2}}\Sigma_k^0 m_\Phi
\eea
The comoving energy density of $S$  produced from $\Phi$ is evaluated from the above equation with the initial condition $\rho_{S}^\Phi (a_{\rm end})=0$.
As mentioned earlier the production of $S$ is dominated at $a_{\rm RH}~({\rm or,~at}~T_{\rm RH})$ \cite{Elahi:2014fsa,Barman:2022qgt} and an approximate solution of $\rho_{S}^{\Phi}$ can be obtained at $a_{\rm RH}(T_{\rm RH})$,
\bea
\rho_{S}^{\Phi}(T_{\rm RH})&\approx& \frac{\sqrt{3} (k+2) \rho _{\rm RH}^{3/2}}{(5 k-8) M_p^3}  \left(\left(\frac{\rho_{\rm end}}{\rho_{\rm RH}}\right)^{\frac{5
   k-8}{6k}}-1\right) \Sigma_k^0 ~m_\Phi,
\eea
Note that starting from an equation with scale factor  $a$ as variable, we obtain the density as a function of $T_{\rm RH}$. The energy density of the SM bath at reheating is given as $\rho_{\rm RH}=\pi^2/30 g_\rho(T_{\rm RH}) T_{\rm RH}^4$, with $g_{\rho}(T)$ being the relativistic degrees of freedom.
Since the $S$ particles have negligible interactions with other particles, once produced, their density only redshifts.
Thus $\rho_{S}^\Phi (T)$ at any other temperature $T(<T_{\rm RH})$ by simple scaling which follows from the conservation of co-moving energy density.
To keep the discussion of this analysis more general and the results applicable to a broader range of
inflationary scenarios, we keep the inflaton mass as a free parameter in our analysis. This is the bare mass term during the reheating. For example,
if we consider a quartic inflationary potential, one needs to add this term to study
reheating around its minimum (see. Refs. \cite{Dimopoulos:2017xox,Ghoshal:2023phi}). 
In certain inflationary scenarios, $T_{RH}$ may be uniquely fixed from the potential, although in the presence of non-minimal coupling, $T_{RH}$ can vary significantly \cite{Clery:2022wib,Barman:2022qgt}. The effect of such non-minimal coupling in this scenario is discussed at the end of this section.
\begin{figure}[!tbh]
    \centering
    \subfigure[\label{trs}]{
     \includegraphics[scale=0.395]{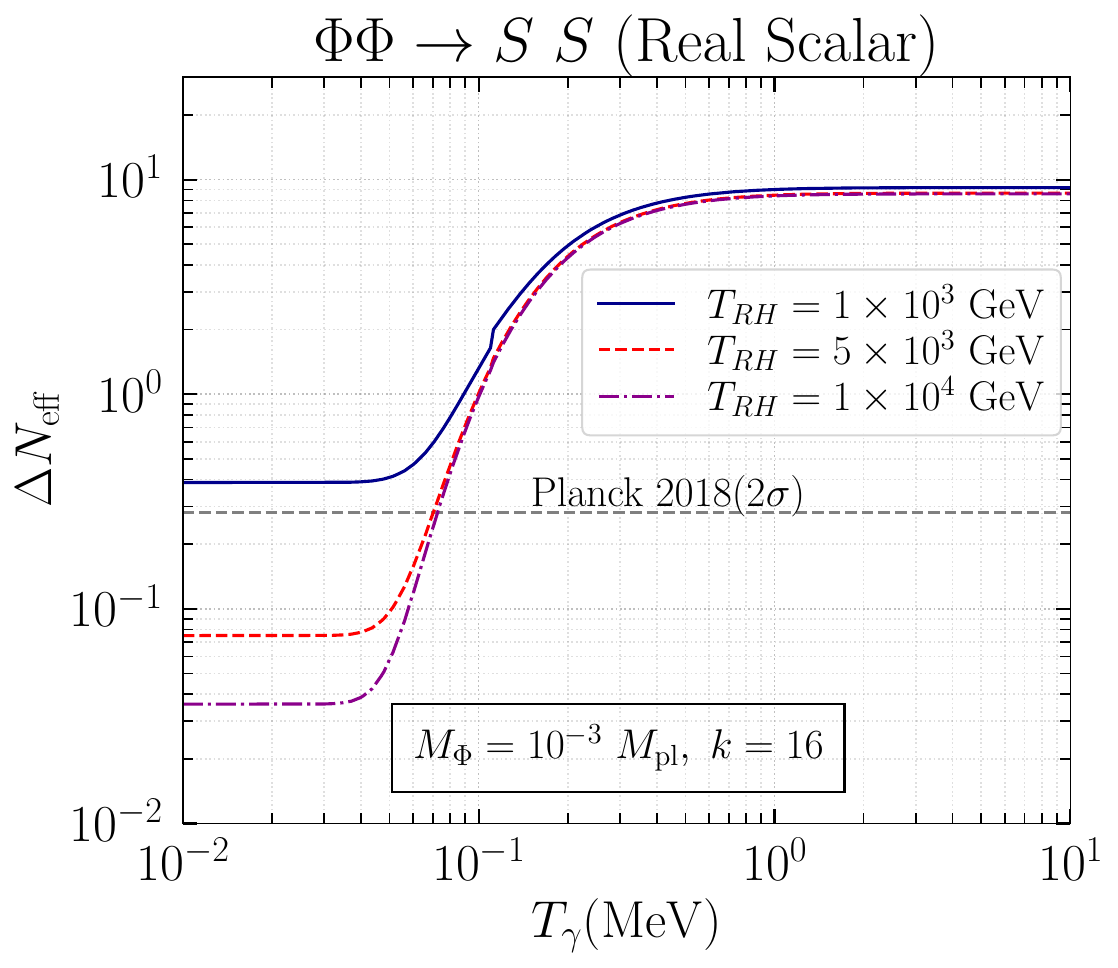}}
    \subfigure[\label{ks}]{
     \includegraphics[scale=0.395]{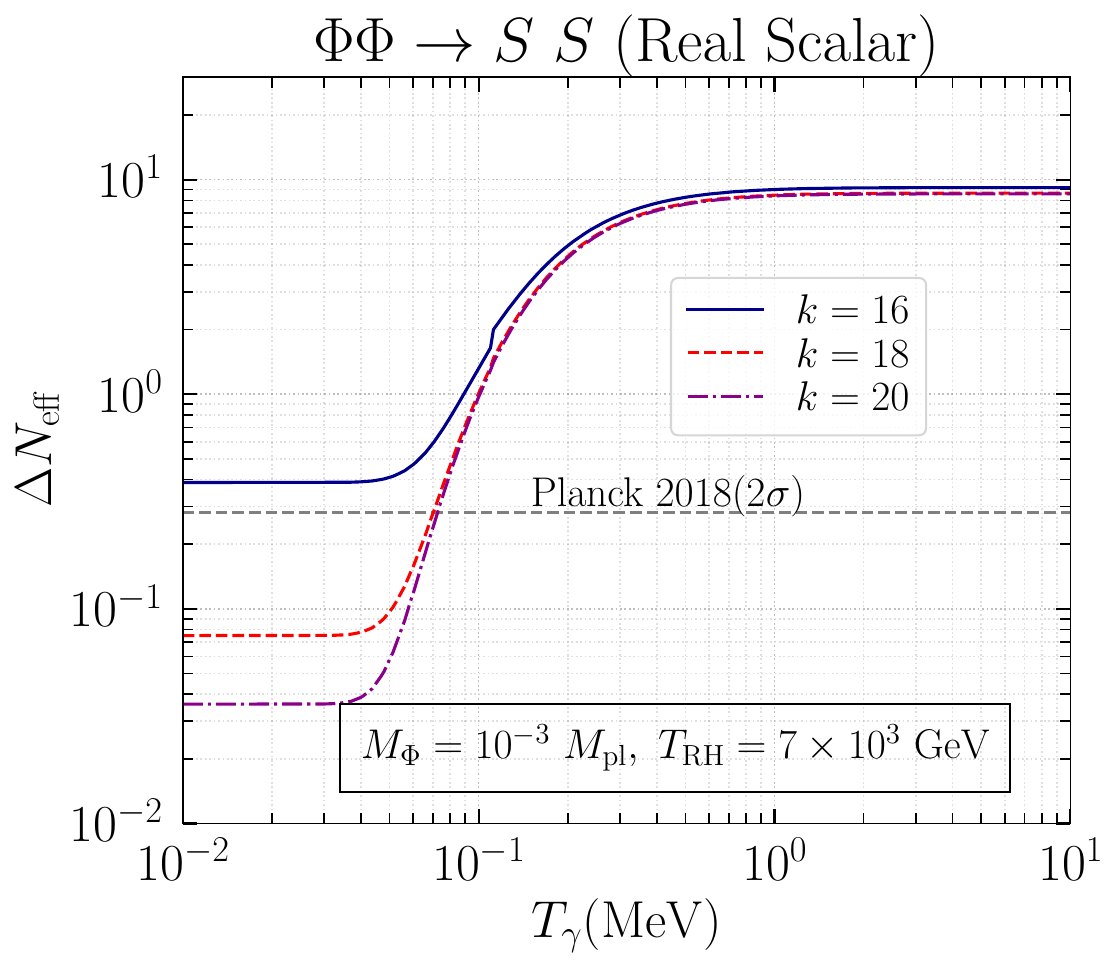}}
    \caption{\it Evolution of $\Delta N_{\rm eff}= N_{\rm eff}-3.046$ with temperature $T_\gamma$ in presence of gravitationally produced scalar radiation with inflaton mass $M_{\Phi}=10^{-3}~M_{\rm pl}$. (a) For different values of $T_{\rm RH}= 10^3~{\rm GeV},~5\times 10^3~{\rm GeV}~{\rm and}~10^4~{\rm GeV}$ shown by blue solid,  red dashed and magenta dashed dot lines respectively with fixed $k=16$.
    (b)For different values of $k=16,~18$ and 20 shown by blue solid,  red dashed and magenta dashed dot lines respectively with fixed $T_{\rm RH} =7\times 10^3$ GeV.}
    \label{fig:neff_vary_scal}
\end{figure}

The contribution in dark scalar density from the gravity-mediated scatterings of SM fields 
(produced gravitationally from inflaton) i.e. $\rho_{S}^{\rm SM}(T)$ can also be obtained from eq.\eqref{eq:rho_SM} as a function of the reheating temperature $T_{\rm RH}$ following Sec.\ref{subsub:sm}.
For $T_{\rm RH}\gtrsim \mathcal{O}(300)$ GeV, where SM particles can be considered as massless,
the coefficient in eq.\eqref{eq:rho_SM} in this scenario becomes $\beta_{S}=0.019$ (obtained from eq.\eqref{eq:csmact}).

After adding up both the contribution we get the total energy density i.e. $\rho_{S}^{\rm tot}(T)=\rho_{S}^{\rm SM}(T)+\rho_{S}^{\Phi}(T)$
and evaluate $N_{\rm eff}$ from eq.\eqref{eq:neff}   following eq.\eqref{eq:tgamma}-eq.\eqref{eq:tx} to track the evolution of energy densities.
In principle, one should track the energy densities starting from $T_{\rm RH}$ until the $\nu_L$ decoupling.
However, in our study, the produced radiation interacts only gravitationally, which is significantly weaker than the expansion rate.
Hence, the produced density only dilutes with time and for simplicity, we can start the Boltzmann equation from $T_\gamma=10$ MeV with the input of $\rho_X(T=10{\rm ~MeV})$ redshifted from $T_{\rm RH}$ to $10$ MeV.

The variation in $\Delta N_{\rm eff}$ with $T_\gamma$ in presence of $S$ is shown in Fig.\ref{fig:neff_vary_scal} for a fixed $M_{\Phi}=10^{-3}~M_{\rm pl}$.
In Fig. \ref{trs} we consider a fixed $k=16$ and showcase the variation in $\Delta N_{\rm eff}= N_{\rm eff}-3.046$ for different values of $T_{\rm RH}= 10^3~{\rm GeV},~5\times 10^3~{\rm GeV}~{\rm and}~10^4~{\rm GeV}$ shown by blue solid ,  red dashed and magenta dashed dot lines respectively.
For our region of interest temperatures we find $\rho^{\rm tot}_S(T)$ is dominated by the contribution from inflaton scattering.
Note that, the red-shifted energy density $\rho^{\Phi}_S(T)$ at temperature $T$ is proportional to $\sim T_{\rm RH}^{\frac{16-4k}{3k}}$.
Hence, $\rho^{\Phi}_S(T)$ has an inverse dependence on $T_{\rm RH}$ for $k>4$ \cite{Barman:2022qgt}.
Consequently, the BSM contribution in extra radiation density affecting the $\nu_L$ decoupling is expected to decrease with an increase in $T_{\rm RH}$.
This feature is prominent in the aforementioned figure, where we observe 
$\Delta N_{\rm eff}$ to be smaller  for higher values of $T_{\rm RH}$.
In Fig. \ref{ks} we show the variation in $\Delta N_{\rm eff}$ for a fixed $T_{\rm RH} =7\times 10^3$ GeV and different values of $k=16,~18$ and 20 shown by blue solid ,  red dashed and magenta dashed dot lines respectively.
The dependence on $k$ can also be understood from the above discussion. 

\begin{figure}[!tbh]
    \centering
    \subfigure[\label{s1}]{
     \includegraphics[scale=0.35]{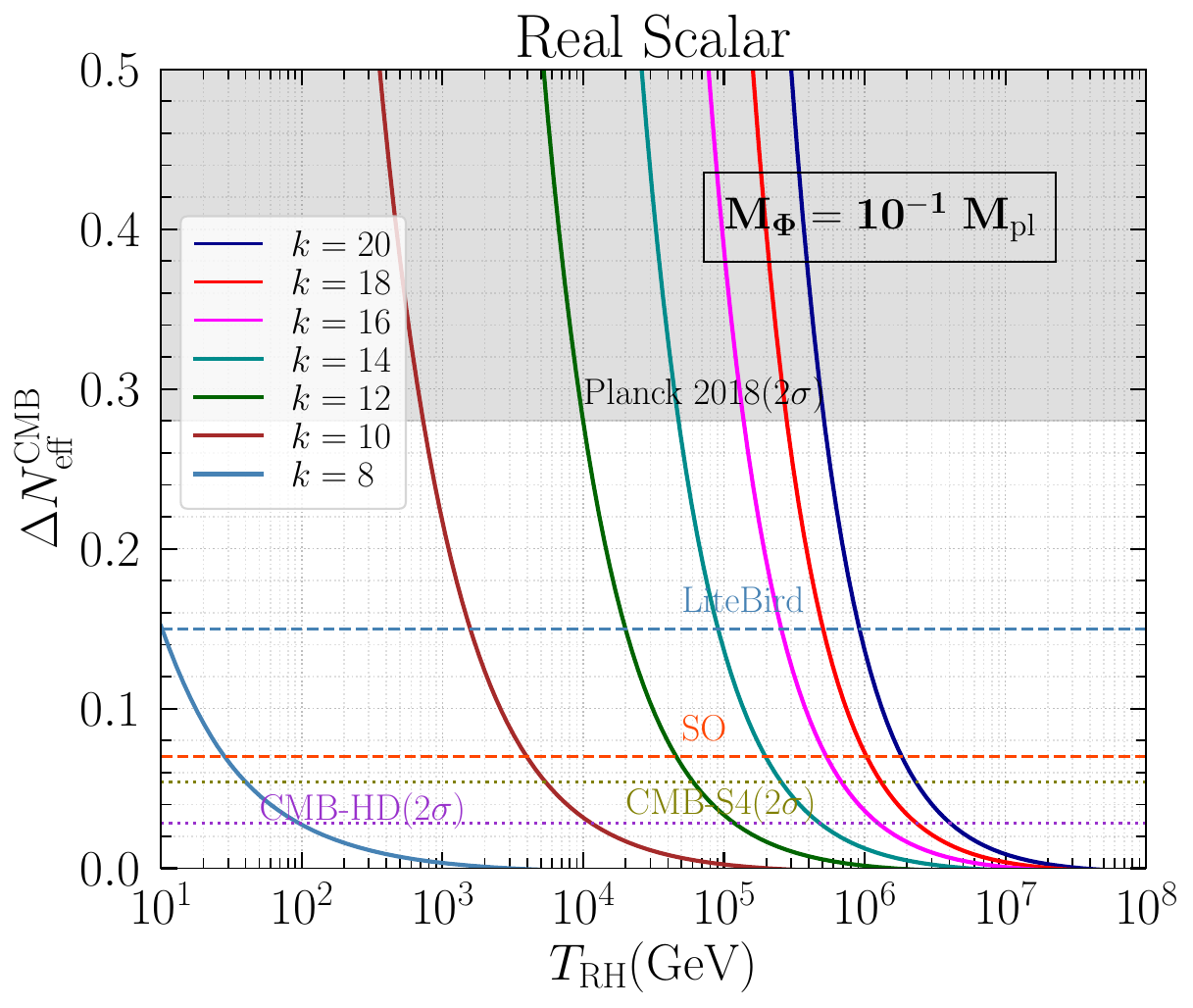}}
    \subfigure[\label{s2}]{
     \includegraphics[scale=0.35]{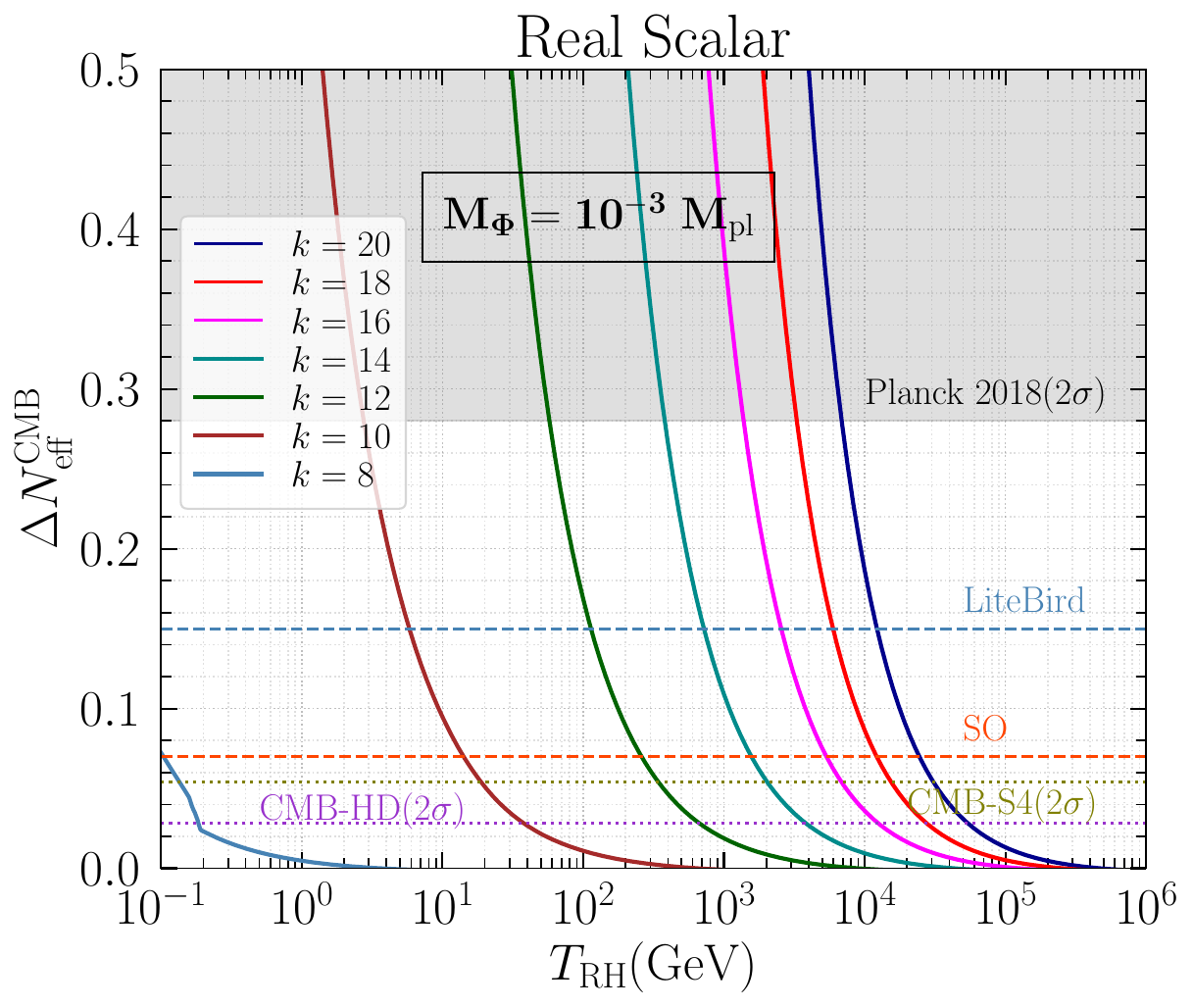}}
     \subfigure[\label{s3}]{
     \includegraphics[scale=0.35]{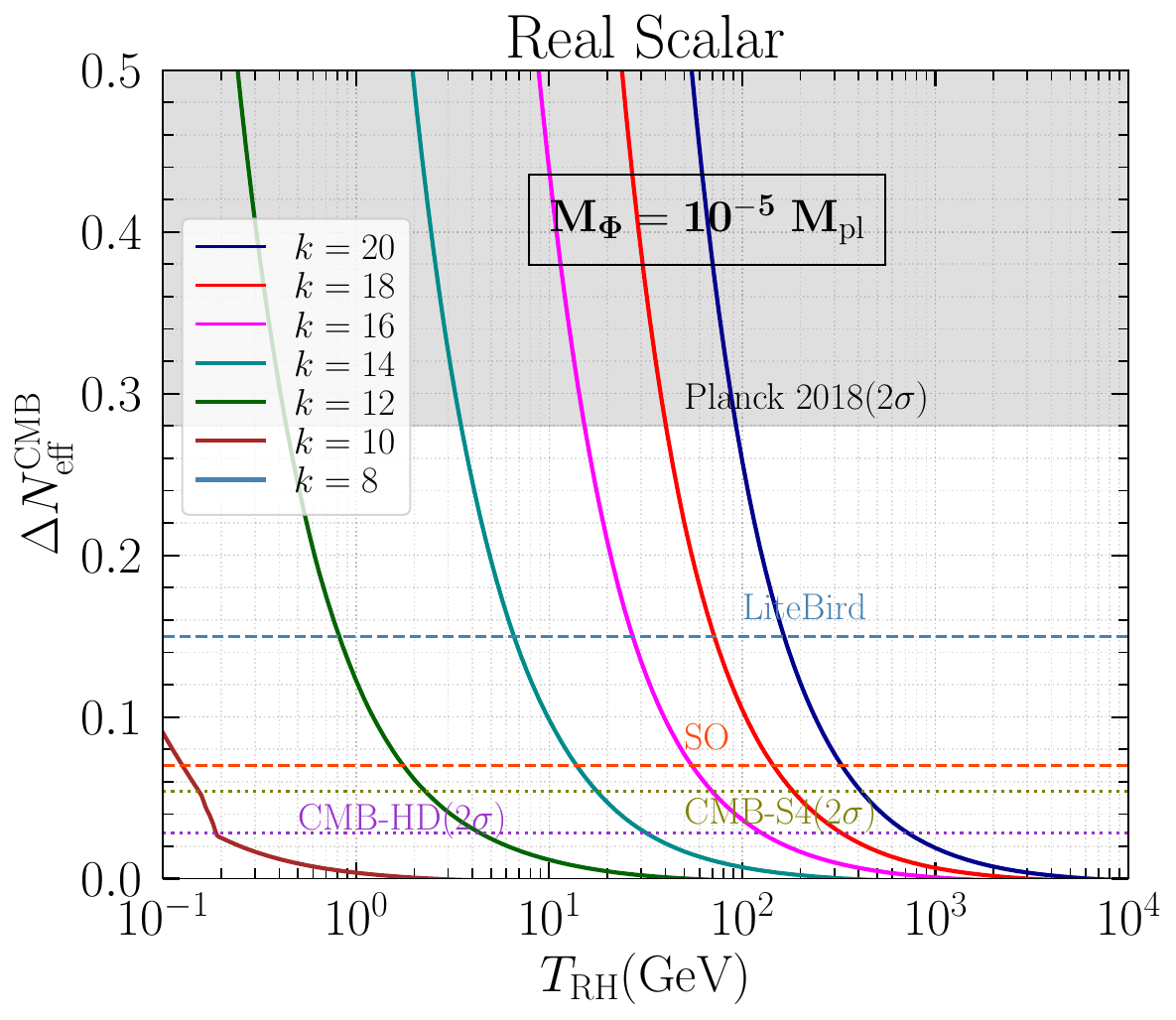}}
    \caption{\it Variation of $\Delta N_{\rm eff}$ with $T_{\rm RH}$  in presence of gravitationally produced scalar radiation for different values of $k\in\{8,~10,~12,~14,~16,~18,~20\}$  depicted by light blue, brown, green, cyan, magenta, red and dark blue solid lines respectively.
    We consider $M_{\Phi}$ to be (a) $10^{-1}~M_{\rm pl}$, (b) $10^{-3}~M_{\rm pl}$ and (c) $10^{-5}~M_{\rm pl}$ in the three panels.
    Existing constraint on $\Delta N_{\rm eff}>0.28$ from Planck 2018 \cite{Planck:2018vyg} at 2$\sigma$ C.L. is depicted by the grey shaded region. Projected future limits from LiteBird  \cite{LiteBIRD:2022cnt}, SO \cite{SimonsObservatory:2019qwx}, CMB-S4 \cite{CMB-S4:2016ple}, CMB-HD \cite{Sehgal:2019ewc} shown by light blue dashed line, orange dashed line, olive dotted lines and magenta dotted lines, respectively. }
    \label{fig:trh_scalar}
\end{figure}

After having a detailed discussion about the dependence of $\Delta N_{\rm eff}$ on the various parameters involved, we now perform a parameter space scan.
In Fig.\ref{fig:trh_scalar} we display the variation of $\Delta N_{\rm eff}$ with $T_{\rm RH}$ for different values of $M_{\Phi}$ and $k$.
We fix $M_{\Phi}$ to be $10^{-1}~M_{\rm pl}$ (Fig.\ref{s1}), $10^{-3}~M_{\rm pl}$ (Fig.\ref{s2}) and $10^{-5}~M_{\rm pl}$ (Fig.\ref{s3}) in the three panels.
The curves of $\Delta N_{\rm eff}$ for different values of $k\in\{8,~10,~12,~14,~16,~18,~20\}$  are signified by light blue, brown, green, cyan, magenta, red and dark blue solid lines respectively. 
For a fixed value of $M_{\Phi}$ and $k$ we observe that $\Delta N_{\rm eff}$ decreases with an increase in $T_{\rm RH}$ as expected from our earlier discussion in context of Fig.\ref{fig:neff_vary_scal}.
Following the same discussion we expect that for same $M_{\Phi}$ and $T_{\rm RH}$. higher values $k$ lead to larger contribution $\Delta N_{\rm eff}$ as  evident in each pannel of Fig.\ref{fig:trh_scalar}.
On the other hand, comparing the three panels we find that for a given value of $k$ and $T_{\rm RH}$,  $\Delta N_{\rm eff}$ is higher for higher values of $M_{\Phi}$.
This can be apprehended from the fact 
the more massive the inflaton field, more energy can be carried by the produced particles.
In the same plane, we also showcase the existing constraint on $\Delta N_{\rm eff}<0.28$ \cite{Planck:2018vyg} from Planck 2018 at 2$\sigma$ C.L. shown by the gray shaded region.
Besides, we also display the constraints from future CMB constraints e.g. LiteBird  \cite{LiteBIRD:2022cnt}, SO \cite{SimonsObservatory:2019qwx}, CMB-S4 \cite{CMB-S4:2016ple}, CMB-HD \cite{Sehgal:2019ewc} shown by light blue dashed line, orange dashed line, olive dotted lines and magenta dotted lines, respectively.
Note that 2$\sigma$ C.L. from Planck 2018 excludes the possibility of light dark scalar radiation even with negligible SM couplings at higher values of $T_{\rm RH}$ depending on $k$ and $M_\Phi$.
For e.g., $T_{\rm RH}> 4\times 10^5$ GeV ($T_{\rm RH}> 8 \times 10^2$ GeV)
is excluded for $k=20~(10)$ for $M_\Phi=10^{-1}~M_{\rm pl}$ (Fig.\ref{s1}).
For $M_\Phi=10^{-5}~M_{\rm pl}$ (Fig.\ref{s3}), $T_{\rm RH}> 9\times 10^1$ GeV ($T_{\rm RH}> 6 \times 10^{-1}$ GeV)
is excluded for $k=20~(12)$.

Future CMB experiments like LiteBird, SO, CMB-S4 and CMB-HD are expected to probe $T_{\rm RH}>  10^6$ GeV ($T_{\rm RH}> 2 \times 10^3$ GeV), 
$T_{\rm RH}>  2 \times 10^6$ GeV ($T_{\rm RH}> 4 \times 10^3$ GeV), $T_{\rm RH}>  3 \times 10^6$ GeV ($T_{\rm RH}> 6 \times 10^3$ GeV) and $T_{\rm RH}>  6 \times 10^6$ GeV ($T_{\rm RH}> 10^4$ GeV) respectively, with $k=20~(10)$ for $M_\Phi=10^{-1}~M_{\rm pl}$ (Fig.\ref{s1}).
For $M_\Phi=10^{-5}~M_{\rm pl}$ (Fig.\ref{s3})
LiteBird, SO, CMB-S4 and CMB-HD should be able to test upto $T_{\rm RH}> 1.5\times 10^2$ GeV ($T_{\rm RH}> 8 \times 10^{-1}$ GeV), $T_{\rm RH}> 3\times 10^2$ GeV ($T_{\rm RH}> 1.8 \times 10^{0}$ GeV), $T_{\rm RH}> 4\times 10^2$ GeV ($T_{\rm RH}> 2 \times 10^{0}$ GeV) and $T_{\rm RH}> 7\times 10^2$ GeV ($T_{\rm RH}> 4 \times 10^{-1}$ GeV)
respectively for $k=20~(12)$.
Since k and $w_{\Phi}$ are related, the constraints and future projections with respect to the background equation of state $w_{\Phi}$ can also be realized following Appendix \ref{apx:a}.
Though we do not show all the future CMB projections on $\nfc$ explicitly, CVL can test the existence of gravity-mediated scalar DR up to
$T_{\rm RH}\sim 10^{9}$ GeV ($T_{\rm RH}\sim 10^{10}$ GeV) and 
$T_{\rm RH}\sim 10^{4}$ GeV ($T_{\rm RH}\sim 10^{7}$ GeV)
for $k=10 (20)$ with $M_\Phi=10^{-1}~M_{\rm pl}$ and $M_\Phi=10^{-5}~M_{\rm pl}$ respectively. This gives us an opportunity to test such feebly interacting light dark higgs which may escape other terrestrial searches.

\subsection*{With non-minimal coupling}
So far, we have presented our analysis assuming minimal coupling with gravity and now we briefly discuss the effect of non-minimal coupling. 
One can consider the same scenario with non-minimal couplings introduced in the scalar sector \cite{Clery:2022wib,Barman:2022qgt},
\begin{equation}
\mathcal{L}_{\rm non-minimal} \supset \xi_\Phi \Phi^2 R+ \xi_H H^2 R, 
\end{equation}
where, $\xi_i$ are the couplings and $H$ is the SM higgs field.
We work in the small field approximation where, the perturbative regime is given by $\xi_i\ll M_{pl}^2/\langle H\rangle^2\sim 10^{33}$ for $\xi_H$( $1$ for $\xi_\Phi$, since $\langle\Phi \rangle\approx M_{pl}$ at the end of inflation \cite{Clery:2022wib}).
After transforming to the Einstein frame, the effective interaction with non-minimal couplings with DR scalar becomes, 
\begin{equation}
 \mathcal{L}_{\rm int}=m_S^2 S^2 \frac{1}{M_{pl}^2}( \xi_\Phi \Phi^2 +  \xi_H H^2) .  
\end{equation}
Thus the production amplitude square for the processes: $HH\to SS$ and $\Phi\Phi\to S S$, due to the non-minimal interaction are given by \cite{Clery:2022wib,Barman:2022qgt},
\bea 
|\mathcal{M}|_{HH\to SS}^2= \bigg( \xi_H \frac{m_{S}^2 }{M_p^2} \bigg)^2,~~~|\mathcal{M}|_{\Phi\Phi\to SS}^2= \bigg( \xi_\Phi \frac{m_{S}^2 }{M_p^2} \bigg)^2
\eea
Substituting the amplitudes in eq.\eqref{eq:cf2} the collision term for production from SM reads as,
\begin{eqnarray}
    \mathcal{C}_{\rm SM}^{\rm non-min.}\propto \xi_{\Phi} ^2 T^5~m_{S}^4/M_{pl}^{4},
\end{eqnarray}
apart from $\mathcal{O}(0.1)$ numbers. 
It is worth pointing out that, compared to the minimal scenario (see eq.\eqref{eq:ctr}), the collision rate for $S$ production from $H$ is suppressed by,
$\sim (m_{S}/T)^4$, apart from the non-minimal coupling.
On the other hand, the production from inflaton scatterings, which provides the dominant contribution as discussed earlier, is suppressed by $(\xi_\Phi m_S)^2$ (see eq.\eqref{eq:cscal}). 
We explicitly check that even with maximal $\xi_{H/\Phi}$ and  $m_{S}$, contribution in $\Delta N_{\rm eff}$ due to the non-minimal interaction is negligible throughout the temperature range from $0.1~{\rm GeV}<T_{\rm RH}<10^{18}$ GeV.
Thus, our analysis of dark Higgs production as  \dr~ along with the cosmological constraints remains similar even with non-minimal couplings in the scalar sector.

\section{Dark Photons as Dark Radiation}
\label{sec:vector}
In a similar fashion to the previous discussion, this section explores the gravity-mediated production of light vector particles as \dr. 
To realize such scenario, we consider a BSM vector field often referred as ``dark photon" with the following {\it Lagrangian} \cite{Caputo:2021eaa},
\bea
\mathcal{L}_{A'}\supset -\frac{1}{4} V_{\mu \nu} V^{\mu \nu} +\frac{1}{2}m_{A'}^2 {A'}^2,
\eea
where, $m_{A'}$ is the mass of this ``dark photon" and $V^{\mu \nu}$ is dark photon field tensor.
Again, we assume that the gauge coupling associated with it as well as the kinetic mixing with SM photon are negligibly small. Henece graviton mediated processes remain as the main channel for $A'$ production. We now analyze the production of $A'$  in the same framework discussed before.

\begin{figure}[!tbh]
    \centering
    \subfigure[\label{trv}]{
     \includegraphics[scale=0.35]{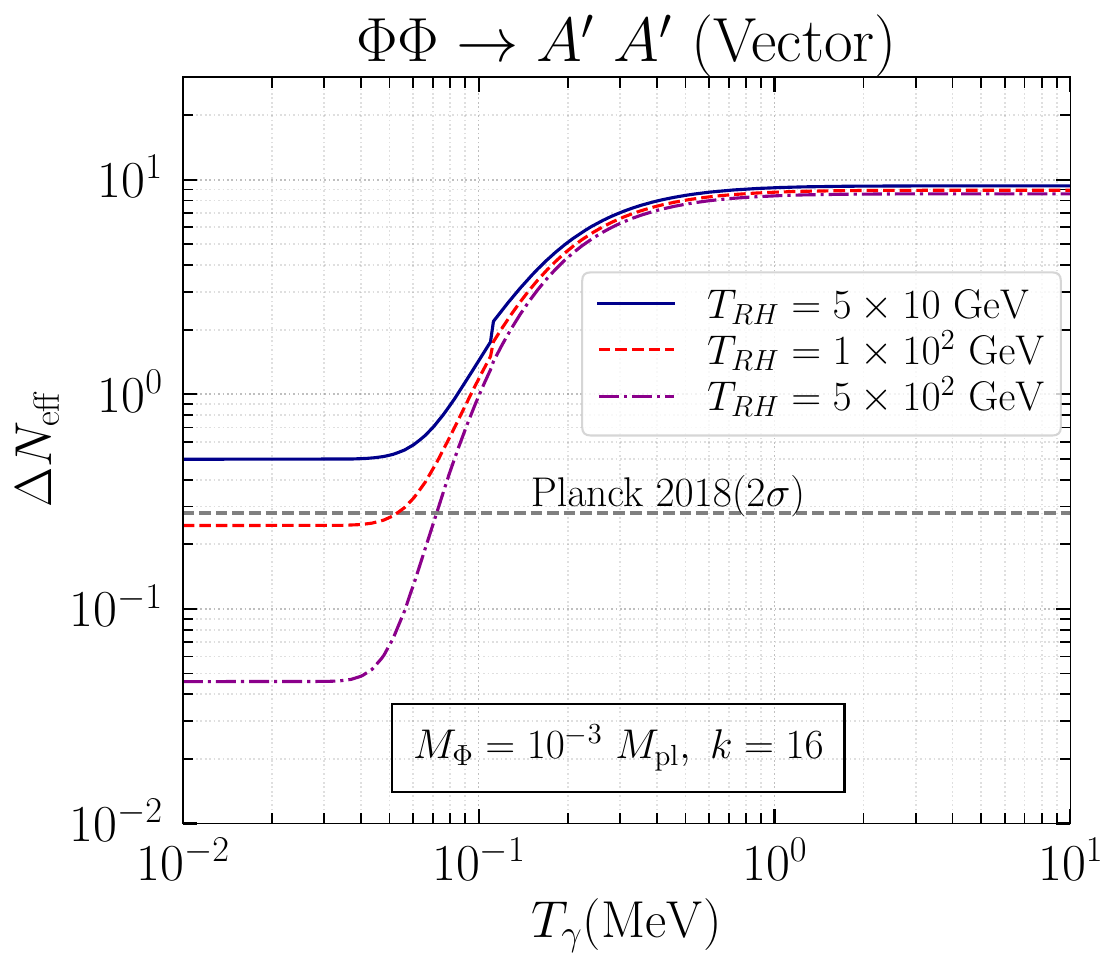}}
    \subfigure[\label{kv}]{
     \includegraphics[scale=0.35]{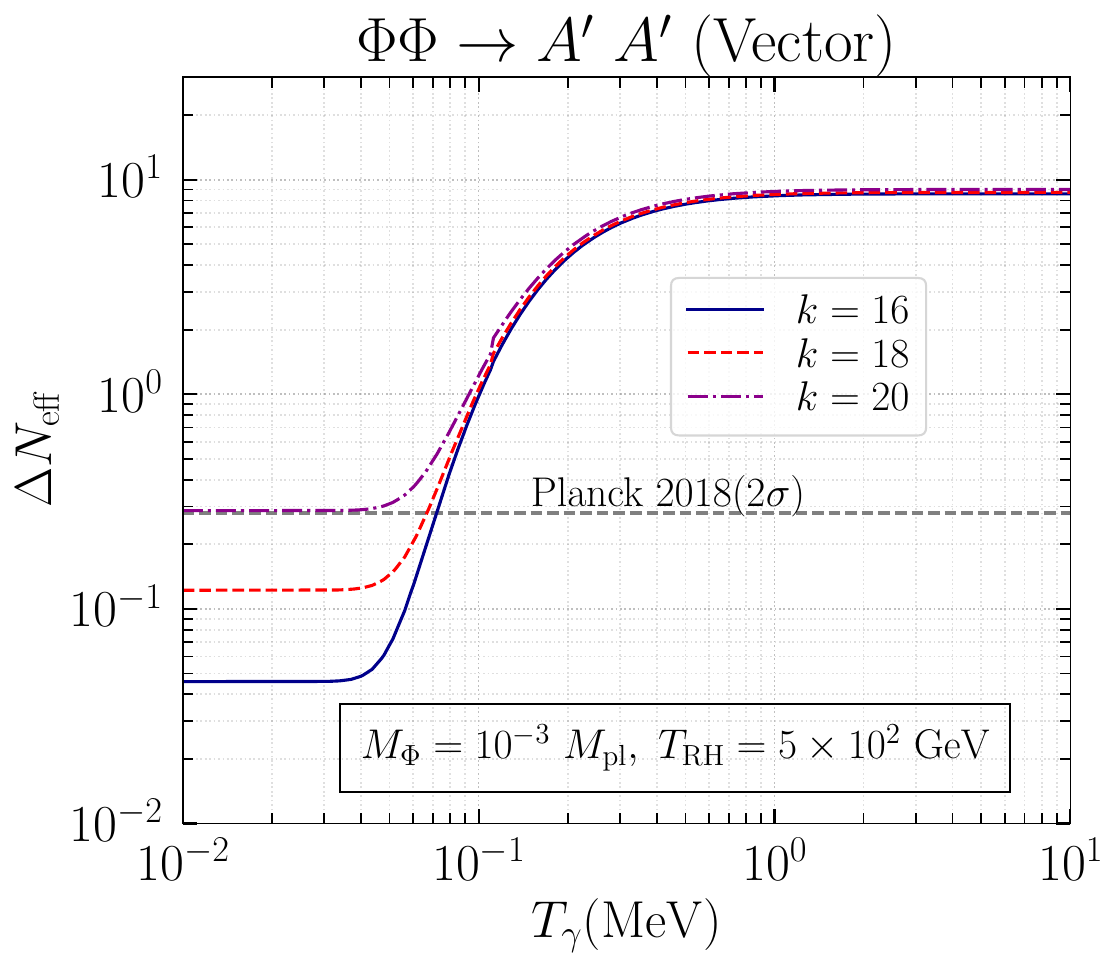}}
    \caption{\it Evolution of $\Delta N_{\rm eff}= N_{\rm eff}-3.046$ with temperature $T_\gamma$ in presence of gravitationally produced vector radiation with inflaton mass $M_{\Phi}=10^{-3}~M_{\rm pl}$. (a) For different values of $T_{\rm RH}= 5\times 10~{\rm GeV},~10^2~{\rm GeV}~{\rm and}~5\times 10^2~{\rm GeV}$ shown by blue solid,  red dashed and magenta dashed dot lines respectively with fixed $k=16$.
    (b)For different values of $k=16,~18$ and 20 shown by blue solid,  red dashed and magenta dashed dot lines respectively with fixed $T_{\rm RH} =5\times 10^2$ GeV.}
    \label{fig:neff_vary_dp}
\end{figure}

We first compute the contribution from inflaton scattering in $A'$ production
following the methodology prescribed in Sec.\ref{subsub:inf}. The relevant collision term in eq.\eqref{eq:cinf}  for the process $\Phi+\Phi \to A'+A'$ can be obtained by substituting the amplitude $|\mathcal{M}|^2_{\Phi \Phi\to A' A'}$ (see ref.\cite{Barman:2021ugy}) which leads to,
\bea
\mathcal{C}_{\Phi\to A'}
&\approx & \frac{g_V}{l!} \frac{m_\Phi \rho_{\Phi}^2}{M_p^4} \Sigma_k^1
~~~{\rm where,} ~\Sigma_k^1 = \frac{1}{8192~\pi } \sum_{n=1} |\mathcal{P}^k_n|^2 
\eea
Again, $l$ signifies the number of identical particles in the final state e.g. $l=2$ for vector particles. $g_V=2$ is the number of produced $V$ in a single reaction. We the substitute the collision term in eq.\eqref{eq:rhoinf} to obtain the energy density evoulution equation,
\bea
\frac{d(\rho_{A'}^{\Phi} a^4)}{da}&=&\frac{\sqrt{3}}{M_p^3} a^3 \rho _{\text{RH}}^{3/2} 
   \left(\frac{a_{\text{RH}}}{a}\right)^{\frac{9 k}{k+2}} \Sigma_k^1 m_\Phi 
\eea
We then follow the same steps as discussed in the previous \ref{sec:scalar} to get the density of $A'$ produced from inflaton ($\rho_{A'}^{\Phi}(T)$) which also has a dependence on $T_{\rm RH}$.
The contribution from SM fields 
produced (gravitationally) from inflaton in generating  $A'$ density  $\rho_{A'}^{\rm SM}(T)$ can be found as a function of the reheating temperature $T_{\rm RH}$    from eq.\eqref{eq:rho_SM}.
In the massless limit, the coefficient in the collision term in eq.\eqref{eq:rho_SM} becomes $\beta_{A'}\sim0.008$  in this scenario.
Again, we sum up both the contributions to achieve, $\rho_{A'}^{\rm tot}(T)=\rho_{A'}^{\rm SM}(T)+\rho_{A'}^{\Phi}(T)$.

\begin{figure}[!tbh]
    \centering
    \subfigure[\label{v1}]{
     \includegraphics[scale=0.35]{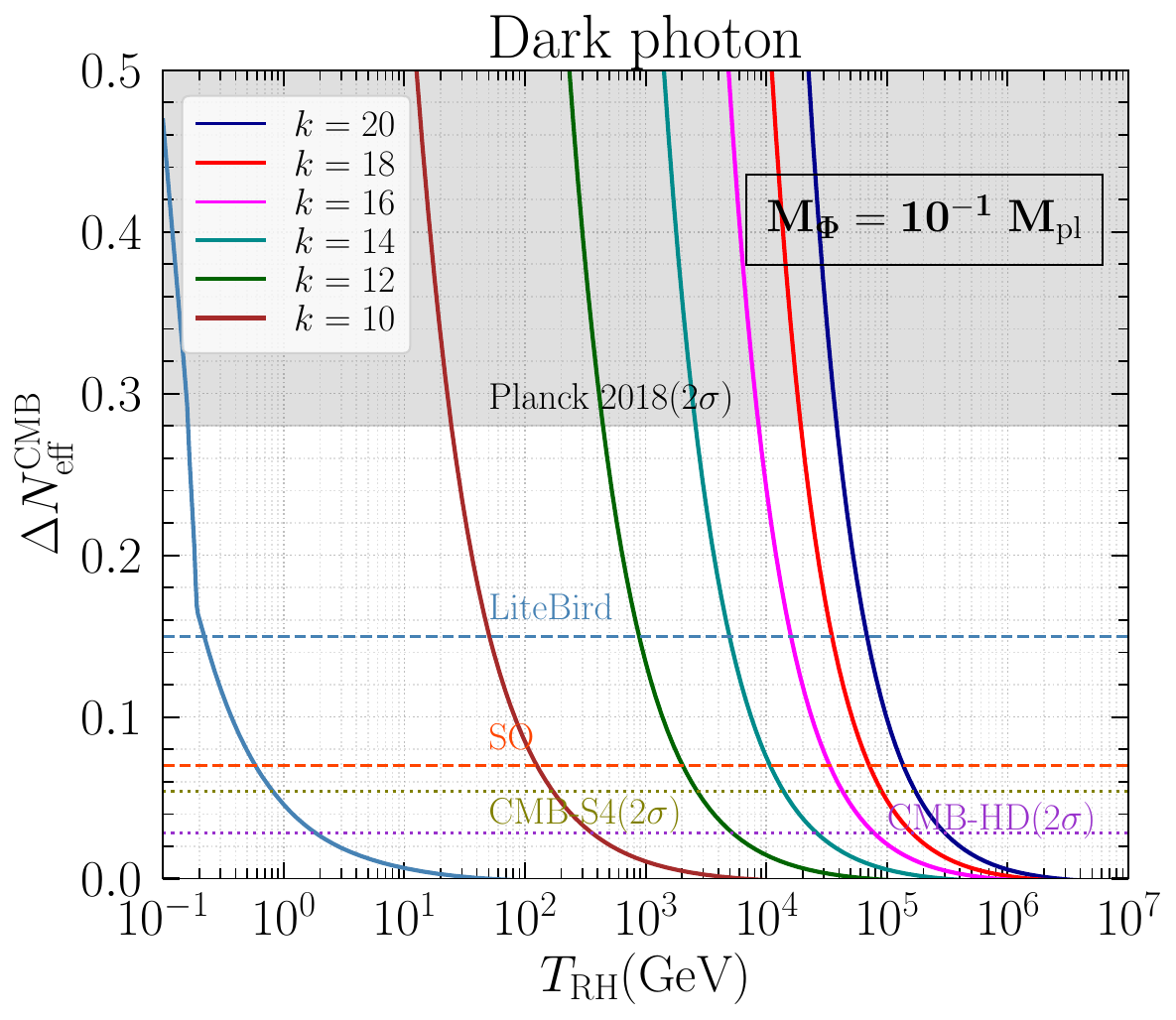}}
    \subfigure[\label{v2}]{
     \includegraphics[scale=0.35]{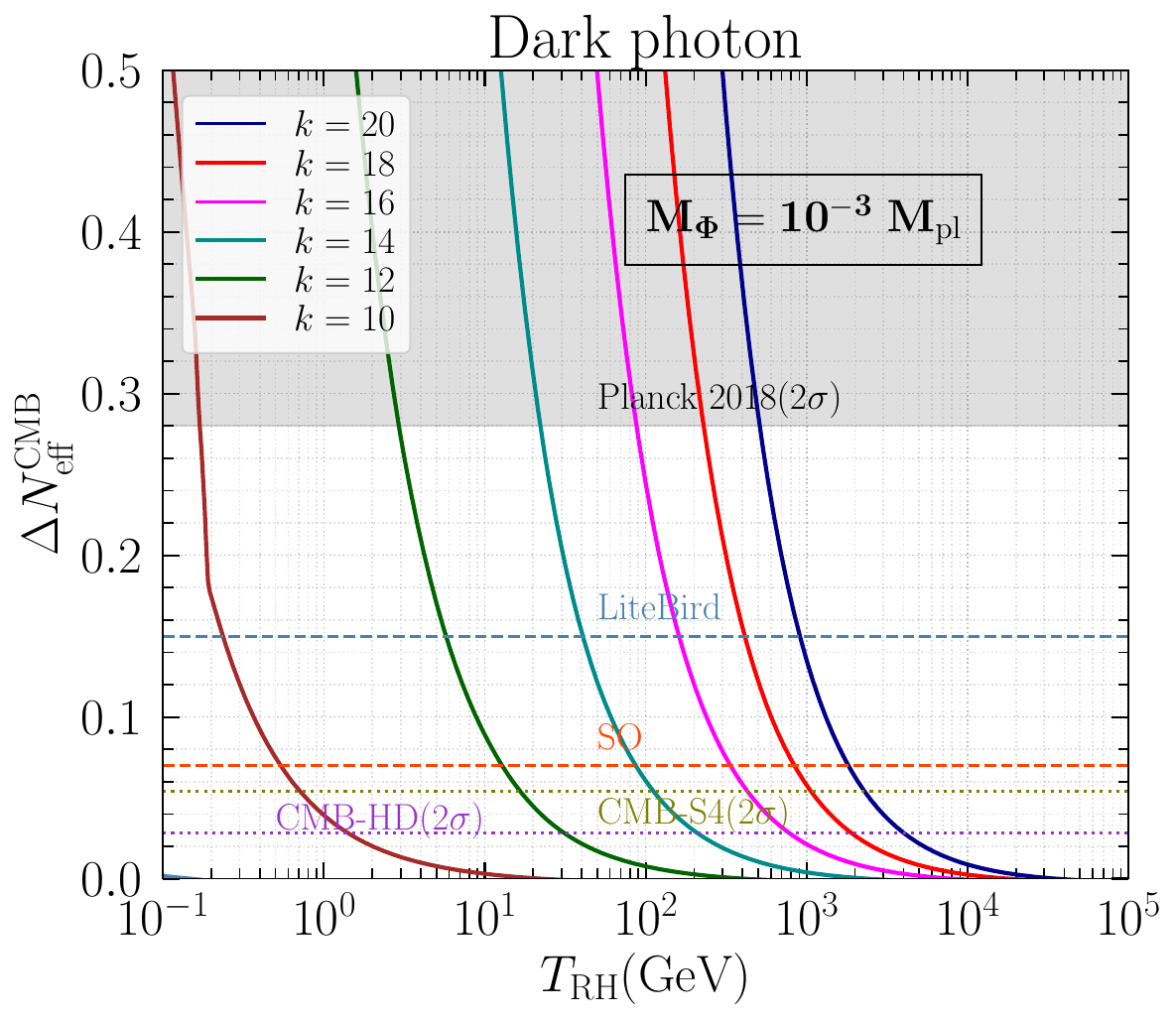}}
     \subfigure[\label{v3}]{
     \includegraphics[scale=0.35]{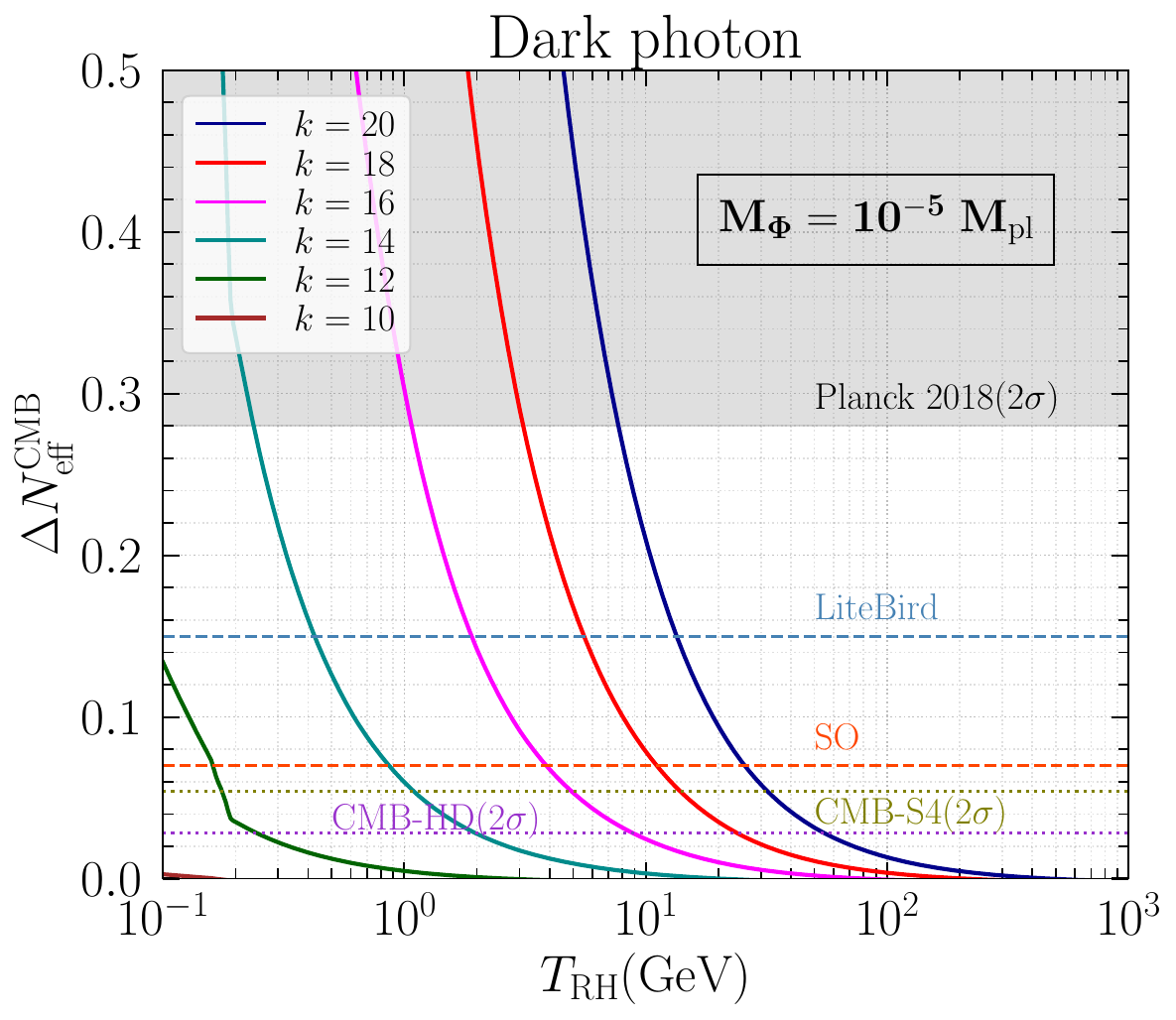}}
    \caption{\it Variation of $\Delta N_{\rm eff}$ with $T_{\rm RH}$  in presence of gravitationally produced vector radiation for different values of $k\in\{8,~10,~12,~14,~16,~18,~20\}$  depicted by light blue, brown, green, cyan, magenta, red and dark blue solid lines respectively.
    We consider $M_{\Phi}$ to be (a) $10^{-1}~M_{\rm pl}$, (b) $10^{-3}~M_{\rm pl}$ and (c) $10^{-5}~M_{\rm pl}$ in the three panels.
     Existing constraint on $\Delta N_{\rm eff}>0.28$ from Planck 2018 \cite{Planck:2018vyg} at 2$\sigma$ C.L. is depicted by the grey shaded region. Projected future limits from LiteBird  \cite{LiteBIRD:2022cnt}, SO \cite{SimonsObservatory:2019qwx}, CMB-S4 \cite{CMB-S4:2016ple}, CMB-HD \cite{Sehgal:2019ewc} shown by light blue dashed line, orange dashed line, olive dotted lines and magenta dotted lines, respectively. }
    \label{fig:trh_dp}
\end{figure}

The variation in $\Delta N_{\rm eff}$ with $T_\gamma$ in presence of $A'$ is shown in Fig.\ref{fig:neff_vary_dp} for a fixed $M_{\Phi}=10^{-3}~M_{\rm pl}$.
In Fig. \ref{trv} we consider a fixed $k=16$ and showcase the variation in $\Delta N_{\rm eff}$ for different values of $T_{\rm RH}= 5\times 10~{\rm GeV},~1\times 10^2~{\rm GeV}~{\rm and}~5\times 10^2~{\rm GeV}$ shown by blue solid ,  red dashed and magenta dashed dot lines respectively.
As discussed in the previous discussion, $\rho^{\Phi}_{A'}(T)$ at temperature $T$  has an inverse dependence on $T_{\rm RH}$ for $k>4$ \cite{Barman:2022qgt} and hence 
 we observe smaller 
$\Delta N_{\rm eff}$  for higher values of $T_{\rm RH}$ for the vector radiation as well.
In Fig. \ref{kv} we show the variation in $\Delta N_{\rm eff}$ for a fixed $T_{\rm RH} =5\times 10^2$ GeV and different values of $k=16,~18$ and 20 shown by blue solid ,  red dashed and magenta dashed dot lines respectively.
The dependence on $k$ can also be understood from the earlier discussion.

In Fig.\ref{fig:trh_dp} we display the variation of $\Delta N_{\rm eff}$ with $T_{\rm RH}$ for different values of $M_{\Phi}$ and $k$ in the presence of light vector radiation.
We fix $M_{\Phi}$ to be $10^{-1}~M_{\rm pl}$ (Fig.\ref{v1}), $10^{-3}~M_{\rm pl}$ (Fig.\ref{v2}) and $10^{-5}~M_{\rm pl}$ (Fig.\ref{v3}) in the three panels.
The curves of $\Delta N_{\rm eff}$ for different values of $k\in\{8,~10,~12,~14,~16,~18,~20\}$  are signified by light blue, brown, green, cyan, magenta, red and dark blue solid lines respectively. 
Our earlier conclusion about the dependence of  $\Delta N_{\rm eff}$ on  $M_\Phi$, $k$ and $T_{\rm RH}$ in context of Fig.\ref{fig:trh_scalar}
hold true for vector radiation too.
Analogus to the scalar case here also we showcase the existing constraint from Planck 2018 at 2$\sigma$ C.L.,  and  projected LiteBird, SO, CMB-S4 and CMB-HD limits at 2$\sigma$ C.L. through the same color combination. 
Note that 2$\sigma$ C.L. from Planck 2018 excludes the possibility of light dark vector radiation even with negligible SM couplings for $T_{\rm RH}> 5\times 10^4$ GeV ($T_{\rm RH}> 2 \times 10^1$ GeV)
with $k=20~(10)$ for $M_\Phi=10^{-1}~M_{\rm pl}$ (Fig.\ref{v1}).
For $M_\Phi=10^{-5}~M_{\rm pl}$ (Fig.\ref{v3}), $T_{\rm RH}> 8$ GeV ($T_{\rm RH}> 3 \times 10^{-1}$ GeV)
is excluded for $k=20~(14)$ in the presence of vector dark radiation.

On the other hand, future CMB experiments like LiteBird, SO, CMB-S4 and CMB-HD are expected to probe $T_{\rm RH}>  8 \times 10^4$ GeV ($T_{\rm RH}> 5 \times 10^1$ GeV), 
$T_{\rm RH}>  1.5 \times 10^5$ GeV ($T_{\rm RH}> 10^2$ GeV), $T_{\rm RH}>  2 \times 10^5$ GeV ($T_{\rm RH}> 2 \times 10^2$ GeV) and $T_{\rm RH}>  3 \times 10^5$ GeV ($T_{\rm RH}> 4 \times 10^2$ GeV) respectively, with $k=20~(10)$ for $M_\Phi=10^{-1}~M_{\rm pl}$ (Fig.\ref{v1}).
Similarly, for $M_\Phi=10^{-5}~M_{\rm pl}$ (Fig.\ref{v3}).
LiteBird, SO, CMB-S4 and CMB-HD should be able to test upto $T_{\rm RH}> 1.1\times 10$ GeV ($T_{\rm RH}> 4 \times 10^{-1}$ GeV), $T_{\rm RH}> 2.5\times 10$ GeV ($T_{\rm RH}> 8 \times 10^{-1}$ GeV), $T_{\rm RH}> 3\times 10$ GeV ($T_{\rm RH}> 1.2 \times 10^{0}$ GeV) and $T_{\rm RH}> 6\times 10$ GeV ($T_{\rm RH}> 2 \times 10^{0}$ GeV)
CVL can test the existence of gravity-mediated vector DR up to
$T_{\rm RH}\sim 10^{8}$ GeV ($T_{\rm RH}\sim 10^{9}$ GeV) and 
$T_{\rm RH}\sim 10^{3}$ GeV ($T_{\rm RH}\sim 5\times 10^{5}$ GeV)
for $k=10 (20)$ with $M_\Phi=10^{-1}~M_{\rm pl}$ and $M_\Phi=10^{-5}~M_{\rm pl}$ respectively.
Of course, the constraints and future projections with respect to the background equation of state $w_{\Phi}$ can also be realized following Appendix \ref{apx:a}.

\section{Generic Spin-2 mediated Dark Radiation Production}
\label{sec:genspin2}
This section is a generalization of previous Sec.\ref{sec:scalar} where we study the dark scalar radiation production in presence of a generic spin-2 mediator with the following interaction Lagrangian \cite{Bernal:2018qlk},
\be
\sqrt{-g}\mathcal{L}_{\rm int}= \frac{1}{2 \Lambda} \Tilde{h}_{\mu \nu}\lt(T^{\mu \nu}_{\rm SM}+ T^{\mu \nu}_{X}+ T^{\mu \nu}_{\Phi}\rt),
\label{eq:spin2}
\ee
where, $\Tilde{h}_{\mu \nu}$ and $\Lambda$ signify the spin-2 field and the effective scale, respectively. 
As mentioned earlier the form of the stress-energy tensor of a field,
$T_s^{\mu \nu}$ depends on its spin $s=0,1/2,1$.
For brevity, we discuss only the scalar dark radiation production i.e. $X=S$.
The mass of the spin-2 mediator ($m_{\Tilde{h}}$) becomes irrelevant when we work in the range $\Lambda>T_{\rm RH}\gg m_{\Tilde{h}}$ \cite{Bernal:2018qlk}. 
For the case when $m_{\Tilde{h}}\gg T_{\rm RH}$ the rates are anyway suppressed by extra $\sim 1/m_{\Tilde{h}}^4$ and hence the  contribution in $\Delta N_{\rm eff}$ is expected to be less significant.
In this work we restrict our analysis in the parameter range $T_{\rm RH}\gg m_{\Tilde{h}}$
and only analyze the scalar \dr~ production in such a scenario.
Analogous to the previously discussed gravity-mediated production (sec.\ref{subsec:prod}), spin-2 mediated $S$ production rate can also be computed by only substituting relevant cross-sections \cite{Bernal:2018qlk}.

\begin{figure}[!tbh]
    \centering
    \subfigure[\label{g1}]{
     \includegraphics[scale=0.35]{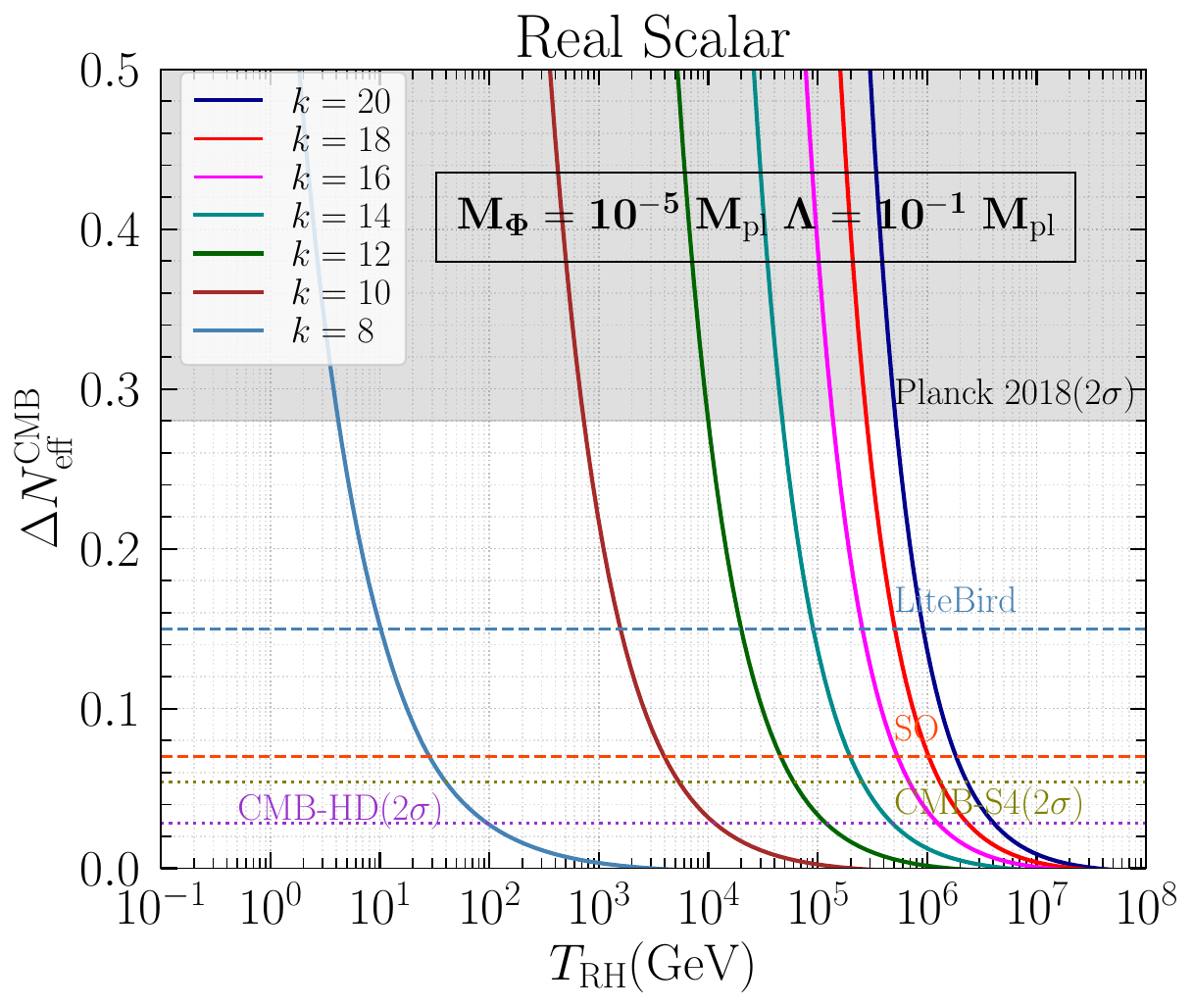}}
    \subfigure[\label{g2}]{
     \includegraphics[scale=0.35]{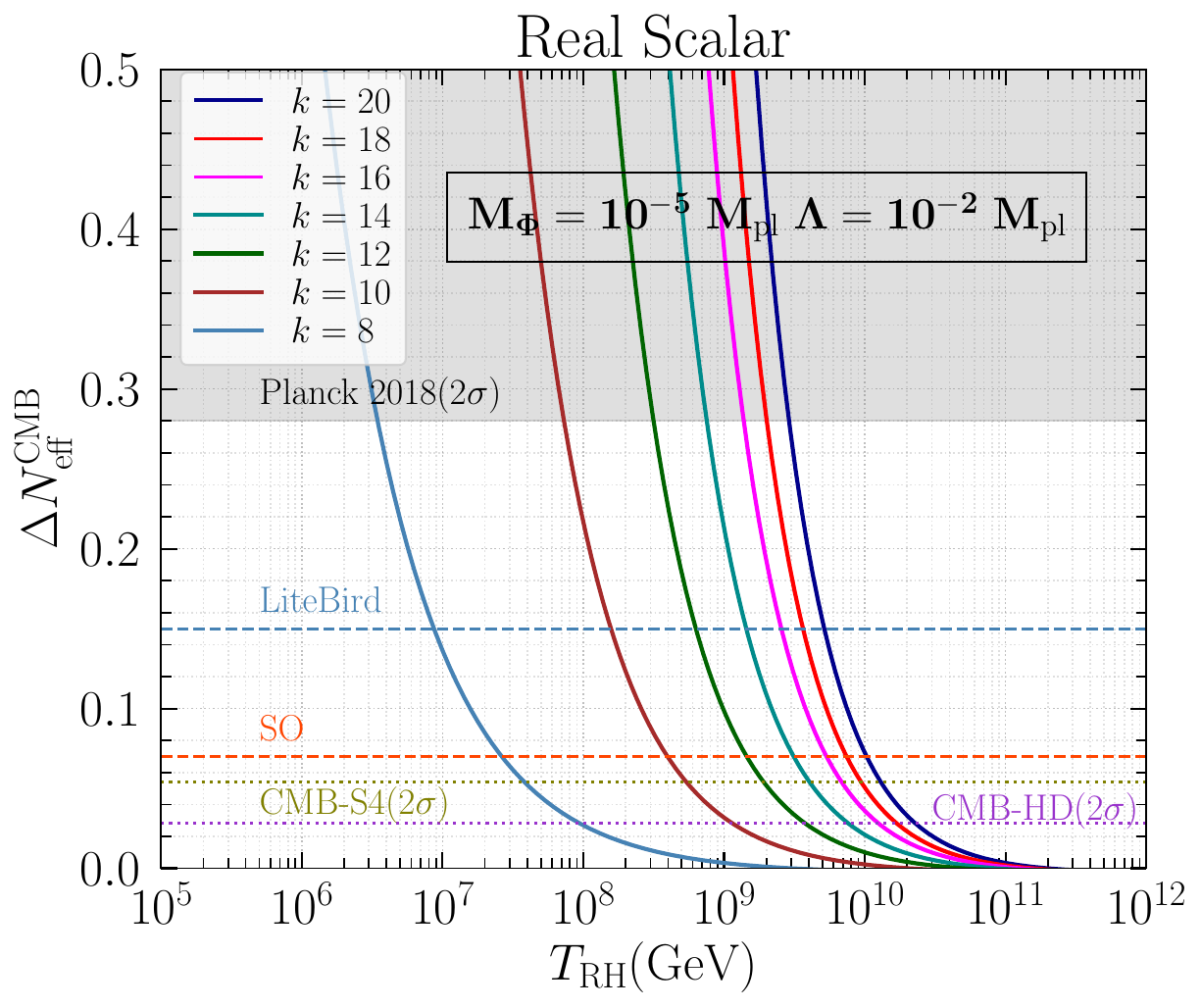}}
    \caption{\it Variation of $\Delta N_{\rm eff}$ with $T_{\rm RH}$ in presence of scalar radiation produced through a spin-2 mediator for different values of $k\in\{8,~10,~12,~14,~16,~18,~20\}$  depicted by light blue, brown, green, cyan, magenta, red and dark blue solid lines respectively.
    We consider $M_{\Phi}$ to be (a) $10^{-1}~M_{\rm pl}$, (b) $10^{-3}~M_{\rm pl}$ and (c) $10^{-5}~M_{\rm pl}$ in the three panels.
     Existing constraint on $\Delta N_{\rm eff}>0.28$ from Planck 2018 \cite{Planck:2018vyg} at 2$\sigma$ C.L. is depicted by the grey shaded region. Projected future limits from LiteBird  \cite{LiteBIRD:2022cnt}, SO \cite{SimonsObservatory:2019qwx}, CMB-S4 \cite{CMB-S4:2016ple}, CMB-HD \cite{Sehgal:2019ewc} shown by light blue dashed line, orange dashed line, olive dotted lines and magenta dotted lines, respectively. }
    \label{fig:trh_spin2}
\end{figure}

In Fig.\ref{fig:trh_spin2} we present  the dependence of $\Delta N_{\rm eff}$ on $T_{\rm RH}$ for different values of $k$ and a fixed $M_{\Phi}=10^{-5}~M_{\rm pl}$.
To signify the $\Delta N_{\rm eff}$ contours of different values of $k\in\{8,10,12,14,16,18,20\}$ and the existing constraints, we use the same color convention as used throughout this paper. 
We consider $\Lambda=10^{-1}~M_{\rm pl}$ and $10^{-2}~M_{\rm pl}$ in 
Fig.\ref{g1} and Fig.\ref{g2} respectively.
For a fixed value of $M_{\Phi}$ and $k$ one can infer that $\Delta N_{\rm eff}$ decreases with an increase in $T_{\rm RH}$, which follows from our earlier discussion in the context of Fig.\ref{fig:neff_vary_scal}.
It is worth highlighting that, for the same set of parameters i.e. $M_{\Phi},k$ and $T_{\rm RH}$ the contribution in $\Delta N_{\rm eff}$ from generic spin 2 particle mediated scattering is significantly stronger than the same with graviton mediated scattering.
The production rate of $S$ is inversely proportional to the associated effective scale ($\sim 1/\Lambda^4$) and the scale chosen here is smaller than that of the graviton mediator ($\sim M_{\rm pl}$).
Comparing the two panels of the aforementioned figure, it can be understood that, with smaller values of $\Lambda$ the contribution in $\Delta N_{\rm eff}$ is larger as expected from the same previous discussion.
Consequently, even larger values of $T_{\rm RH}$ is excluded from the spin-2 mediated $S$ production compared to the graviton only case.
For example, 2$\sigma$ C.L. from Planck 2018 excludes the possibility of light scalar radiation even with negligible SM couplings for $T_{\rm RH}> 5\times 10^6$ GeV ($T_{\rm RH}> 8 \times 10^2$ GeV)
with $k=20~(10)$ for $\Lambda=10^{-1}~M_{\rm pl}$ (Fig.\ref{g1}).
For $\Lambda=10^{-2}~M_{\rm pl}$ (Fig.\ref{g2}), $T_{\rm RH}> 4\times 10^9$ GeV ($T_{\rm RH}> 3\times 10^7$ GeV)
is excluded for $k=20~(10)$ in the presence of scalar dark radiation.

Future CMB experiments like LiteBird, SO, CMB-S4 and CMB-HD are expected to probe $T_{\rm RH}>  6 \times 10^9$ GeV ($T_{\rm RH}> 1.1 \times 10^8$ GeV), 
$T_{\rm RH}>   10^{10}$ GeV ($T_{\rm RH}> 4 \times 10^8$ GeV), $T_{\rm RH}>  1.1 \times 10^{10}$ GeV ($T_{\rm RH}> 6 \times 10^8$ GeV) and $T_{\rm RH}>  2 \times 10^{10}$ GeV ($T_{\rm RH}>  10^9$ GeV) respectively, with $k=20~(10)$ for $\Lambda=10^{-2}~M_{\rm pl}$ (Fig.\ref{g2}).
Similarly, for $\Lambda=10^{-1}~M_{\rm pl}$  (Fig.\ref{g1})
LiteBird, SO, CMB-S4 and CMB-HD should be able to test upto $T_{\rm RH}> 10^6$ GeV ($T_{\rm RH}> 1.2 \times 10^{3}$ GeV), $T_{\rm RH}> 2\times 10^6$ GeV ($T_{\rm RH}> 4 \times 10^{3}$ GeV), $T_{\rm RH}> 3\times 10^6$ GeV ($T_{\rm RH}> 6 \times 10^{3}$ GeV) and $T_{\rm RH}> 4\times 10^6$ GeV ($T_{\rm RH}> 10^{4}$ GeV).
CVL can test the existence of spin 2-mediated scalar DR up to
$T_{\rm RH}\sim 10^{9}$ GeV ($T_{\rm RH}\sim 7 \times 10^{9}$ GeV) and 
$T_{\rm RH}\sim 10^{12}$ GeV ($T_{\rm RH}\sim 5\times 10^{13}$ GeV)
for $k=10 (20)$ with $\Lambda=10^{-1}~M_{\rm pl}$  and $\Lambda=10^{-2}~M_{\rm pl}$ respectively.
As mentioned earlier the constraints and future projections can be presented also with respect to the background equation of state $w_{\Phi}$ following Appendix \ref{apx:a}.

\begin{figure}[!tbh]
    \centering
    \subfigure[\label{1h}]{
     \includegraphics[scale=0.35]{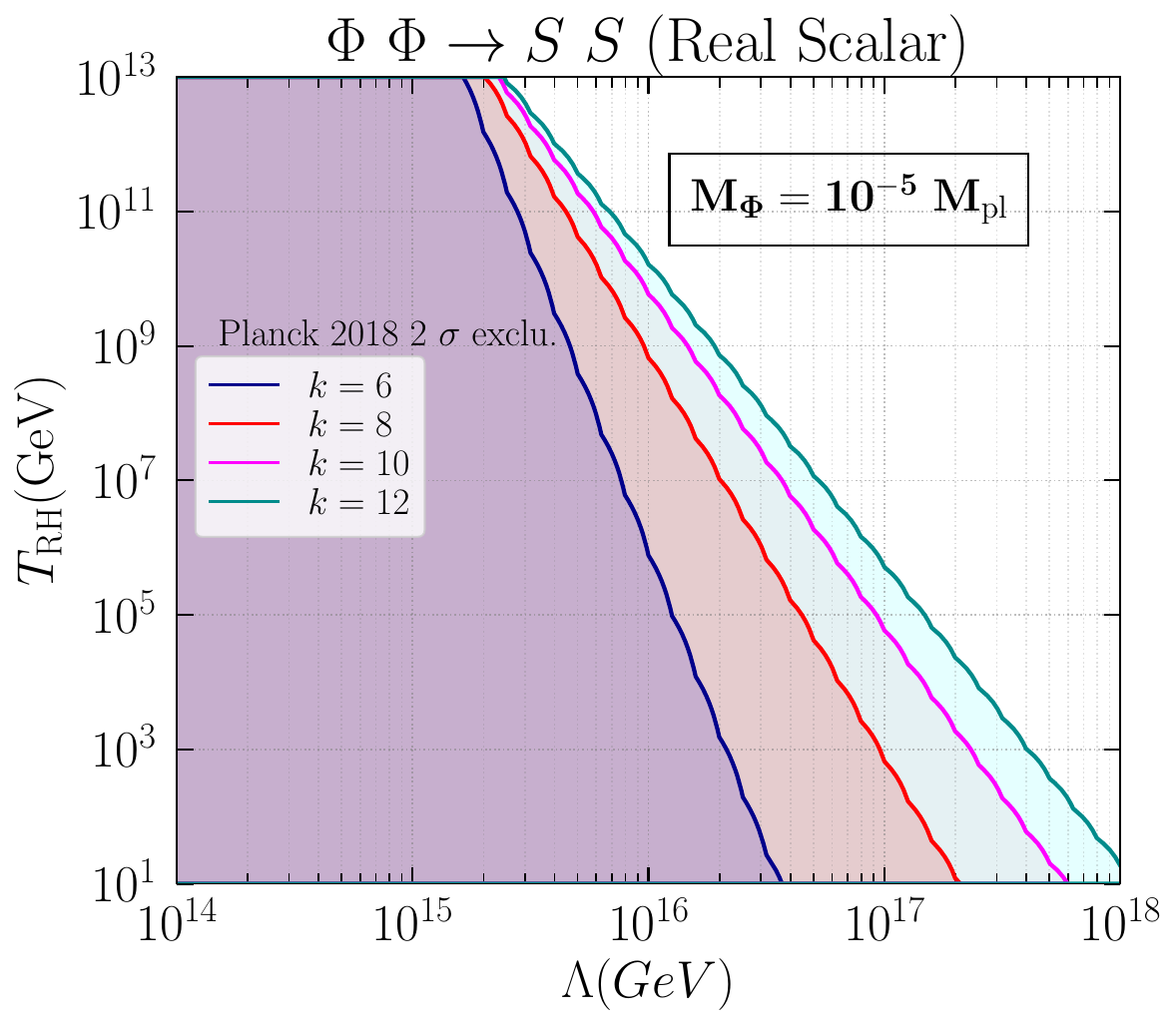}}
    \subfigure[\label{2h}]{
     \includegraphics[scale=0.35]{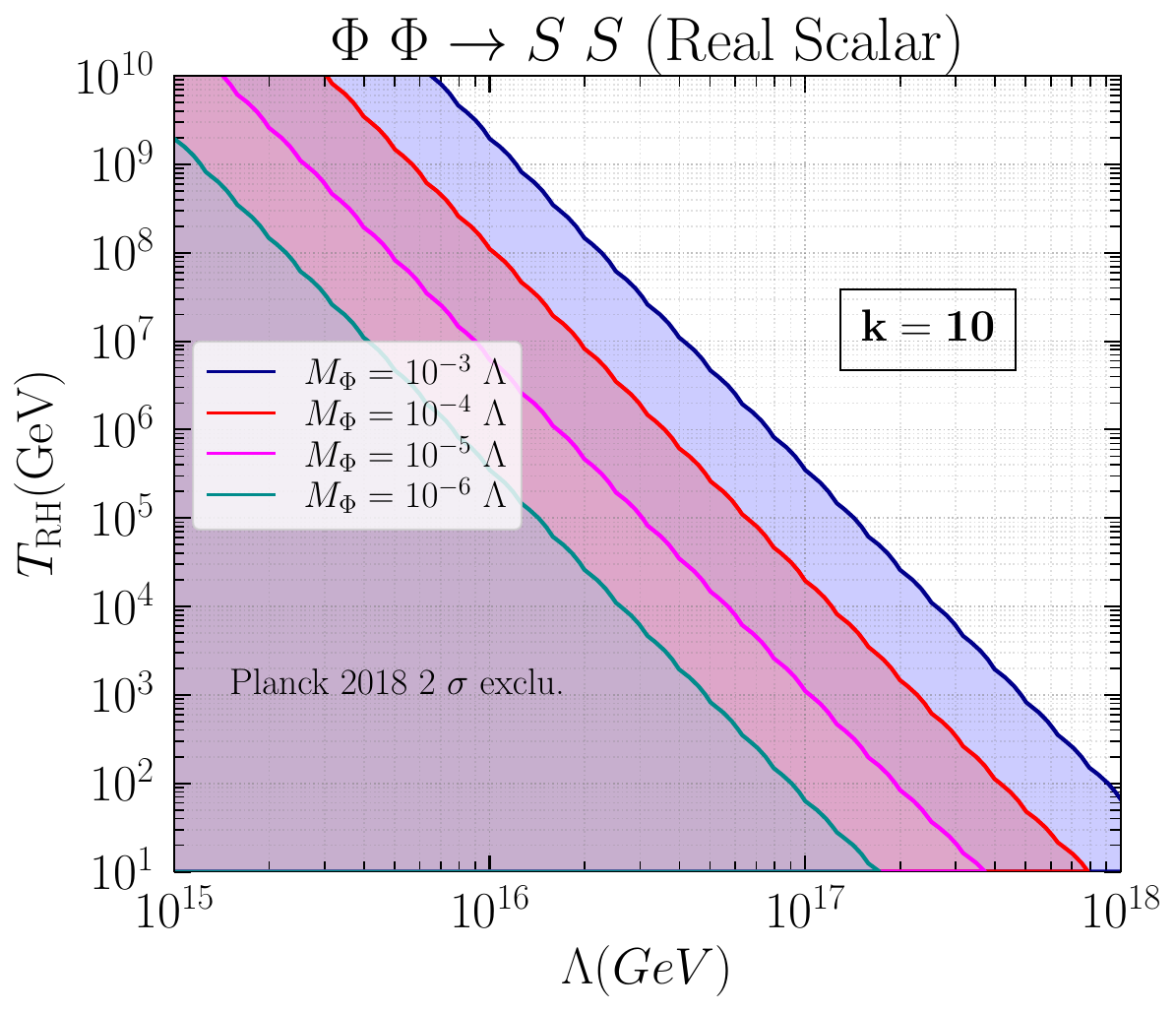}}
    \caption{\it Allowed parameter space from the Planck 2018 $N_{\rm eff}$ constraint at $2\sigma$ limit in the $\Lambda$ vs. $T_{\rm RH}$ plane, considering spin-2 mediated dark scalar production.
    (a) Exclusion regions corresponding to $k=6,~8,~10$ and $12$ are depicted by blue, red, magenta and cyan shaded regions, respectively for fixed inflaton mass $M_\Phi=10^{-5}~M_{\rm pl}$.
    (b) The exclusion regions corresponding to different  $M_\Phi\in(10^{-3}\Lambda,~10^{-4}\Lambda,~10^{-5}\Lambda$ and $10^{-6}\Lambda)$ are depicted by blue, red , magenta and cyan shaded regions For a fixed value of $k=10$.}
    \label{fig:trh_lam_spin2}
\end{figure}

Finally, in Fig.\ref{fig:trh_lam_spin2} we display the allowed parameter space from the Planck 2018 $N_{\rm eff}$ constraint at $2\sigma$ limit in the $\Lambda$ vs. $T_{\rm RH}$ plane, considering spin-2 mediated dark scalar production.
In Fig.\ref{1h} we show the constrained parameter space for some benchmark values of $k$ and a fixed inflaton mass $M_\Phi=10^{-5}~M_{\rm pl}$ which is also the typical mass in $\alpha-$ attractor inflationary models \cite{Barman:2021ugy,Barman:2022qgt,Clery:2021bwz}.
The exclusion regions corresponding to $k=6,~8,~10$ and $12$ are depicted by blue, red, magenta and cyan shaded regions, respectively. 
For example, with $\Lambda=10^{16}$ GeV the only allowed range from Planck 2018 is $T_{\rm RH}\gtrsim10^6$ GeV in such a scenario featuring scalar \dr.  
It is worth highlighting that in the presence of a generic spin-2 mediated scalar \dr~ production with an effective scale $\Lambda\lesssim 10^{17}$ GeV, Planck-2018 sets the strongest upper limit on $T_{\rm RH}$ even for the typical inflationary models.
Note that, for a fixed value of $k$, with an increase in the value of $\Lambda$, the contribution in $\Delta N_{\rm eff}$ decreases due to the suppression of $\sim 1/\Lambda^4$.
On the other hand, $\Delta N_{\rm eff}$ has an inverse dependence on $T_{\rm RH}$ for a fixed value of $\Lambda$ as already observed from Fig.\ref{fig:trh_spin2}.
These two competing effects shape the $2\sigma$ exclusion contours from $\Delta N_{\rm eff}$ to be left-tilted lines in the aforementioned plot.
In Fig.\ref{2h} we display the contours of Planck 2018  $2\sigma$  constraint on $N_{\rm eff}$ for a fixed value of $k=10$ with varying $M_{\Phi}$.
The exclusion regions correspond to $M_\Phi=10^{-3}\Lambda,~10^{-4}\Lambda,~10^{-5}\Lambda$ and $10^{-6}\Lambda$, and are depicted by blue, red, magenta, and cyan shaded regions.
For a fixed value of $k$, with a decrease in the value of $M_\Phi$, the contribution in $\Delta N_{\rm eff}$  also decreases as a result of the smaller amount of energy transferred to radiation from $\Phi$, which is also evident from the same plot.
Thus, with future CMB experiments, the presence of spin-2 mediated dark radiation can be detected, and null observations will set stringent constraints on inflationary parameters.

\section{ALP and Dirac RHN}
\label{sec:ALP}
Apart from the aforementioned scenarios other BSM particles can also be produced as DR from gravity mediated processes. In this context, we highlight two other class of particles (1) fermion DR ($s=1/2$) e.g. Dirac right handed neutrions (RHN) $\nu_R$ and (2) pseudoscalar DR ($s=0$) e.g. Axion-like particles (ALP) $a$.

\begin{enumerate}
    \item For spin 1/2 particle, i.e. the production of $\nu_R$ from inflaton scattering ($\Phi\Phi\to \nu_R \nu_R$) is suppressed by mass of $\nu_R$ ($\propto m_{\nu_R}^2$) \cite{Barman:2021ugy}.
As already mentioned in Sec.\ref{sec:scalar}-Sec.\ref{sec:genspin2}, the DR production for $T_{\rm RH}\lesssim10^{18}$ GeV is dominated by inflaton scattering.
Hence for $\nu_R$ radiation with mass $m_{\nu_R}\sim 0.1$ eV, the produced density is also suppressed, and the contribution in $\nfc$ found is not significant compared to scalar or vector radiation. 

\item On the other hand, ALPs are realized with a potential $V(a)=f_a^2m_a^2(T)(1-\cos{(a/f_a)})$ which can be expanded in the large scale approximation around $a/f_a\approx0$ leading to $V(a)\approx \frac{1}{2} m_a^2(T) a^2 +\mathcal{O}(1/f_a^2)$ \cite{DiLuzio:2020wdo}.
Thus the ALP-graviton coupling will be similar to scalar radiation case at leading order since gravity couples to stress energy tensor which is indeed similar to scalar one (eq.\eqref{eq:sclag}). Hence, the expected constraint from $\nfc$ will be analogous to a scalar that can be deduced from Sec.\ref{sec:scalar}. 

\end{enumerate}

\section{Discussion and Conclusion}
\label{sec:concl}
In this work, we systematically analyzed the production of dark radiation through gravity-mediated processes and its imprint on $N_{\rm eff}$ at the time of the CMB epoch. $\nfc$ has already placed stringent constraints various BSM scenarios featuring any light BSM particles which may affect the radiation energy density. However, despite the feeble direct couplings with SM fields such particles can be inevitably produced in the early universe through gravity-mediated process during the reheating era.
We categorically analyze such production through inflaton scattering and SM particle scattering during reheating by s-channel graviton exchange.
Once produced the BSM particle density is diluted only through the expansion effect since their other SM couplings are negligible, showing up as extra dark radiation energy density at CMB.
In Sec.\ref{sec:prod} we discuss the detailed methodology to evaluate $\nfc$ in the presence of generic dark radiation $X$ produced through gravity-mediated processes (Fig.\ref{fig:fd}). 
Here, we would like to highlight the key findings as follows.

\begin{itemize}
    \item 
    Since the production of dark sector radiation particle $X$ occurs through dimensionful interaction featuring amplitude squares proportional to $s^4/M_{pl}^4$, it depends on the reheating temperature $T_{\rm RH}$ \cite{Elahi:2014fsa}. We derived the relevant collision terms 
    eq.\eqref{eq:cinf} and eq.\eqref{eq:cSM}.

    \item 
    For dark sector scalar S particle, we find that for $T_{\rm RH} \lesssim M_{pl}$, production of $S$ is dominated by $\Phi\Phi\to S S$ scattering. We find that $\Delta N_{\rm eff}$ decreases with an increase in $T_{\rm RH}$ and increases with $M_\Phi$ (see Fig.\ref{fig:trh_scalar}). Utilizing the current data from Planck and other CMB probes involving $\Delta N_{\rm eff}$ we constrain the reheat temperature.
    For e.g., $T_{\rm RH}> 4\times 10^5$ GeV ($T_{\rm RH}> 8 \times 10^2$ GeV) is excluded for $k=20~(10)$ for $M_\Phi=10^{-1}~M_{\rm pl}$ (Fig.\ref{s1}). For $M_\Phi=10^{-5}~M_{\rm pl}$ (Fig.\ref{s3}), $T_{\rm RH}> 8\times 10^2$ GeV ($T_{\rm RH}> 8 \times 10^{-1}$ GeV) is excluded for $k=20~(12)$  2$\sigma$ C.L. from Planck 2018.

    \item
    For dark sector vector boson particle $A'$ 
    we find that 2$\sigma$ C.L. limit on $\Delta N_{\rm eff}$ from Planck 2018 excludes the possibility of light dark vector radiation even with negligible SM couplings for $T_{\rm RH}> 5\times 10^4$ GeV ($T_{\rm RH}> 2 \times 10^1$ GeV)
    with $k=20~(10)$ for $M_\Phi=10^{-1}~M_{\rm pl}$ (Fig.\ref{s1}).
     For $M_\Phi=10^{-5}~M_{\rm pl}$ (Fig.\ref{s3}), $T_{\rm RH}> 8$ GeV ($T_{\rm RH}> 3 \times 10^{-1}$ GeV) is excluded for $k=20~(14)$ in the presence of vector dark radiation (see Fig.\ref{fig:trh_dp} ).
    \item
    For the production of \dr  through a generic spin-2 portal with effective scale $\Lambda<M_{ pl} > T_{\rm RH} $ we find
   for example, significantly large contribution (see Fig.\ref{fig:trh_spin2}) due to the production rate being inversely proportional to the associated effective scale ($\sim 1/\Lambda^4$) (see eq.\eqref{eq:spin2}).
    For example, for $k=10$ and $M_\Phi=10^{-5}~M_{pl}$
    the presence of spin 2 mediator with $\Lambda=10^{17}$ GeV is constrained for $T_{\rm RH}\gtrsim 10^5$ GeV (see Fig.\ref{fig:trh_lam_spin2}) for scalar Dark Radiation production .

    \item Apart from the existing constraints on $\nfc$ we also highlight the future projection on it from next generation CMB experiments like LiteBird ($\nfc\lesssim 3.19$) \cite{LiteBIRD:2022cnt}, Simons Observatory ($\nfc\lesssim 3.12$ at $95\%$ C.L.) \cite{SimonsObservatory:2019qwx}, 
    CMB-S4 ($\nfc\lesssim 3.10$ at $95\%$ C.L.) \cite{CMB-S4:2016ple},
     CMB-HD ($\nfc\lesssim3.06$) \cite{Sehgal:2019ewc}.
     Such experiments are expected to probe even smaller values of $\Delta N_{\rm eff}$ at CMB and thus offer an ideal opportunity to probe gravity mediated dark radiation production in the early universe. Null observations from such experiments will exclude even larger values of $T_{RH}$ for the same set of parameters than the existing constraints.  
     For example, CMB HD will be able probe
     $T_{\rm RH}>  6 \times 10^6$ GeV ($T_{\rm RH}> 10^4$ GeV) respectively, with $k=20~(10)$ for $M_\Phi=10^{-1}~M_{\rm pl}$ (Fig.\ref{s1}) for scalar dark radiation.
     For vector dark radiation, CMB HD will be sensitive to $T_{\rm RH}>  3 \times 10^5$ GeV ($T_{\rm RH}> 4 \times 10^2$ GeV) respectively, with $k=20~(10)$ for $M_\Phi=10^{-1}~M_{\rm pl}$ (Fig.\ref{v1}).

     \item  For scalar dark radiation production in presence of a generic spin-2 mediator, CMB HD will provide the strongest constraint
     $T_{\rm RH}\sim 10^{12}$ GeV ($T_{\rm RH}\sim 5\times 10^{13}$ GeV)
for $k=10 (20)$ with $\Lambda=10^{-1}~M_{\rm pl}$  and $\Lambda=10^{-2}~M_{\rm pl}$ respectively (see Fig.\ref{fig:trh_spin2}) even with $M_\Phi=10^{-5} M_{pl}$. 
\end{itemize}

Though we present our analysis and constraints along with future projections with respect to $k$, the same can be easily translated for the background equation of state $w_{\Phi}$ following Appendix \ref{apx:a}.
For simplicity throughout this work, we assume the BSM particles pose negligible couplings with SM particles. Thus, the produced DR density only undergoes a dilution effect due to the expansion of the universe.
A more involved scenario in which DR particles interact with SM particles may involve additional number-changing interactions apart from the gravitational ones that come to play in setting the densities \cite{Barman:2021ugy}, potentially alter the contributions to $\nfc$.
Self-interacting dark radiation produced through gravity mediated process is another viable scenario that can be interesting to investigate in the future with additional constraints arising due to CMB and LSS \cite{Jeong:2013eza,Das:2025asx}.

Thus, gravity-mediated production is an unavoidable source for BSM radiation even with tiny coupling with SM particles.
Future experiments like Lite-BIRD, CMB-Bharat, Euclid, Simon's Observatory will be extremely sensitive to $\Delta N_{\rm eff}^{\rm CMB}$ and will be probe dark sector scenarios, which is otherwise challenging to probe in traditional laboratory experiments.

\section*{Acknowledgment}
We thank Basabendu Barman for very useful comments on the manuscript.
We thank Md Riajul Haque, Sourav Mondal, and Sourav Pal for helpful discussions.
The work of SJ is supported by the National Natural Science
Foundation of China (12425506 and 12375101) and State Key Laboratory of Dark Matter Physics.

\appendix
\section{Appendix: Reheating Background Equation of State $w_{\Phi}$}
\label{apx:a}

Here we show the equation of state ($w_\Phi\approx \frac{k-2}{k+2}$) \cite{Garcia:2020wiy} for various analyses done in Sec.\ref{sec:scalar}, Sec.\ref{sec:vector}, Sec.\ref{sec:genspin2} in order to understand the results in terms of the effective equation of state $w_{\Phi}$. The limit $w_{\Phi} \rightarrow 1$ for very large k values represents what is known as the most stiff equation of state or ``kination-domination epoch" which have significant impact on detectable Primordial Gravitational Waves \cite{Gouttenoire:2021jhk, Chen:2024roo, Ghoshal:2025emq}.

\begin{table}[H]
    \centering
    \begin{tabular}{|c|c|}
    \hline
        $k$ & $w_{\Phi}$  \\
    \hline
    \hline
        4&  ${1}/{3}$\\
        \hline
        6 & ${1}/{2}$\\
        \hline
         $8$ & ${3}/{5}$\\
         \hline
         10 &2/3\\
         \hline
         12 & 5/7\\
          \hline
         14 & 3/4\\
          \hline
          16 & 7/9\\
           \hline
           18 & 4/5\\
           \hline
           20 & 9/11\\
            \hline
    \end{tabular}
    \caption{\it Correspondence between $k$ and $w_{\Phi}$.}
    \label{tab:inf_amps}
\end{table}

\medskip

\bibliography{ref}
\bibliographystyle{JHEP}

\end{document}